\documentclass[aps,prd,twocolumn]{revtex4}
\usepackage{amsmath,amssymb,amsfonts}
\usepackage{graphicx}
\usepackage{enumerate} 
\usepackage{colordvi} 
\usepackage{bm}
\usepackage{color}

\newcommand{\be}{\begin{eqnarray}}
\newcommand{\ee}{\end{eqnarray}}
\newcommand{\simgt}{\lower.5ex\hbox{$\; \buildrel > \over \sim \;$}}
\newcommand{\simlt}{\lower.5ex\hbox{$\; \buildrel < \over \sim \;$}}

\newcommand{\bfx}{\boldsymbol{x}}
\newcommand{\bfv}{\boldsymbol{v}}

\newcommand{\bfk}{\boldsymbol{k}}

\newcommand{\bfq}{{\bf q}}
\newcommand{\bfe}{{\bf e}}
\newcommand{\bfu}{{\bf u}}
\newcommand{\bfr}{{\bf r}}
\newcommand{\bfs}{\boldsymbol{s}}

\newcommand{\tdelta}{\tilde{\delta}}
\newcommand{\dd}{\mathrm{d}}
\newcommand{\ii}{\mathrm{i}}

\begin{document}
\title{Redshift-space equal-time angular-averaged consistency relations of the gravitational dynamics}
\vfill
\author{Takahiro Nishimichi$^{1,2}$, Patrick Valageas$^{3}$}
\bigskip
\affiliation{$^1$Sorbonne Universit\'es, UPMC Univ Paris 06, UMR 7095, Institut d'Astrophysique de Paris, F-75014, Paris, France}
\affiliation{$^2$CNRS, UMR 7095, Institut d'Astrophysique de Paris, F-75014, Paris, France}
\affiliation{$^3$Institut de Physique Th\'eorique, Universit\'e Paris Saclay, CEA, CNRS, F-91191 Gif-sur-Yvette, France}
\bigskip
\date{\today}
%

\begin{abstract}
We present the redshift-space generalization of the equal-time angular-averaged 
consistency relations between $(\ell+n)$- and $n$-point polyspectra of the cosmological 
matter density field. Focusing on the case of $\ell=1$ large-scale mode and $n$ small-scale 
modes, we use an approximate symmetry of the gravitational dynamics to derive explicit 
expressions that hold beyond the perturbative regime, including both the large-scale
Kaiser effect and the small-scale fingers-of-god effects.
We explicitly check these relations, both perturbatively, for the lowest-order version that
applies to the bispectrum, and nonperturbatively, for all orders but for the 
one-dimensional dynamics.
Using a large ensemble of $N$-body simulations, we find that our squeezed bispectrum 
relation is valid to better than $20\%$ up to $1h$Mpc$^{-1}$, for both the monopole 
and quadrupole at $z=0.35$, in a $\Lambda$CDM cosmology. 
Additional simulations done for the Einstein-de Sitter background suggest that these 
discrepancies mainly come from the breakdown of the approximate symmetry of the 
gravitational dynamics. 
For practical applications, we introduce a simple ansatz to estimate the new derivative 
terms in the relation using only observables.
Although the relation holds worse after using this ansatz, we can still recover it 
within $20\%$ up to $1h$Mpc$^{-1}$, at $z=0.35$ for the monopole.
On larger scales, $k = 0.2 h\mathrm{Mpc}^{-1}$, it still holds within the statistical accuracy 
of idealized simulations of volume $\sim8h^{-3}\mathrm{Gpc}^3$ without shot-noise error. 
\end{abstract}

\maketitle

\maketitle
\flushbottom
\section{Introduction}

Accurate understanding of the nonlinear gravitational dynamics is a key for observational 
projects that measure the statistical properties of the cosmic structures on large scales.
The typical scales of interest in these projects range from the weakly to strongly nonlinear
regimes \cite{Albrecht2006,Laureijs2011}.
While perturbation theory is expected to be applicable as long as the nonlinear corrections 
are subdominant \cite{Goroff1986,Bernardeau2002}, a fully nonlinear description would
be helpful to extract cosmological information out of the measured statistics over a wider
dynamic range.
The analytical description also becomes more complicated when one models higher-order 
statistics. Although an increasing number of analytical techniques to calculate the power 
spectrum or the two-point correlation function have been proposed, based for instance on 
resummations of perturbative series expansion or effective approaches
\cite{Crocce2006a,Valageas2007,Pietroni2008,Bernardeau2008,Taruya2012,Crocce2012,Bernardeau2013,Valageas2013,Pietroni2012,Baumann2012,Carrasco2014,Baldauf2014}, 
few of them have been applied to the bispectrum or even higher orders.

Consistency relations between different polyspectra are then very useful to have an 
accurate description of the higher-order statistics once one has a reliable model for the 
lowest-order one, the power spectrum. Alternatively, they can be used to test analytical
models, numerical simulations, or the underlying cosmological scenario (e.g., the impact
of modified gravity or complex dark energy models).
Based on the assumption of Gaussian initial conditions and gravitational dynamics 
governed by general relativity, these relations hold at the nonperturbative level and 
provide a rare insight into the nonlinear regime of gravitational clustering.
 
The most generic consistency relations are ``kinematic consistency relations''
that relate the $(\ell+n)$-density correlation, with $\ell$ large-scale wave numbers
and $n$ small-scale wave numbers, to the $n$-point small-scale density correlation,
with $\ell$ prefactors that involve the linear power spectrum at the
large-scale wave numbers \cite{Kehagias2013,Peloso2013,Creminelli2013,Kehagias2014,Peloso2014,Creminelli2014,Valageas:2014aa,Creminelli2014a,Kehagias2015}.
These relations, obtained at the leading order over the large-scale wave numbers $k_j'$,
arise from the equivalence principle, which ensures that small-scale structures respond
to a large-scale perturbation (which at leading order corresponds to a constant 
gravitational force over the extent of the small-size object) by a uniform displacement.
Therefore, these relations express a kinematic effect, due to the displacement of 
small-scale structures between different times.
This also means that (at this order) they vanish for equal-time statistics, as
a uniform displacement has no impact on the statistical properties of the density
field observed at a given time.
Because they derive from the equivalence principle these relations are very general and
also apply to baryons and galaxies. However, in a standard cosmology they provide no
information at equal times (apart from constraining possible deviations from Gaussian 
initial conditions and general relativity).

To obtain non-vanishing results for equal-time statistics, one must go beyond this
kinematic effect. This implies studying the response of small-scale structures to
non-uniform gravitational forces, which at leading order and after averaging over angles
correspond to a large-scale gravitational curvature.
As proposed in \cite{Valageas:2014ab} and \cite{Kehagias2014a}, this is possible by 
using an approximate symmetry of the gravitational dynamics
(associated with the common approximation $\Omega_\mathrm{m}/f^2  \simeq 1$, 
where $\Omega_\mathrm{m}$ is the matter density cosmological parameter and 
$f=\mathrm{d}\ln D_+/\mathrm{d}\ln a$ is the linear growth-rate), 
which allows one to absorb the change of cosmological parameters
(hence of background curvature) by a change of variable.
These relations again connect the $(\ell+n)$-point polyspectra, with $\ell$ large-scale 
modes and $n$ small-scale modes to the $n$-point polyspectrum, when an angular 
averaging operation is taken over the $\ell$ large-scale modes, which also removes the 
kinematic effect. 
These consistency relations no longer vanish at equal times but they are less general
than the previous relations. Indeed, galaxy formation processes (cooling, star formation,..)
introduce new characteristic scales that would explicitly break this symmetry.
The lowest-order relation, which applies to the matter bispectrum, has been explicitly
tested in \cite{Nishimichi2014} using a large ensemble of cosmological
$N$-body simulations, see also \cite{Chiang2014,Ben-Dayan2015,Wagner2015}
for related discussions and comparisons with simulations or halo models.

The aim of this paper is to generalize this analysis, presented in \cite{Valageas:2014ab} 
and \cite{Nishimichi2014} in real space, to redshift space, where actual observations
take place. This is only a first step towards a comparison with measures from galaxy 
surveys, because we do not consider the important issue of galaxy bias in this paper
(i.e., to translate our results in terms of the galaxy distribution one would need to
add a model that relates the galaxy and matter density fields).
However, this remains a useful task as redshift-space statistics are well-known to be
difficult to model because small-scale nonperturbative effects have a non-negligible
impact up to rather large scales \cite{Scoccimarro1999,Bernardeau2002,Scoccimarro2004},
for instance through the fingers-of-god effect \cite{Jackson1972}.
Therefore, it is even more important than for real-space statistics to build tools
that hold beyond the perturbative regime.

This paper is organized as follows. First, in Sec.~\ref{sec:density} we introduce the 
statistics of the redshift-space density field and its response to the initial conditions. 
Then, in Sec.~\ref{sec:approx-sym} we describe the dynamical equations of the system 
that we consider here and show their symmetry that is valid under the approximation 
$\Omega_\mathrm{m}/f^2 \simeq 1$.
Using these results, we finally derive the angular-averaged consistency relations in 
Sec.~\ref{sec:angular}.
We focus on the lowest-order version of these relations, i.e. the bispectrum, 
in Sec.~\ref{sec:bspectrum}, where we present our results in terms of the multipole 
moments of the spectra. We also introduce a simple ansatz to estimate new derivative 
terms in the relation from observables, to simplify its form and facilitate the
connection with practical situations.
In Sec.~\ref{sec:checks}, the consistency relations are checked both perturbatively and 
nonperturbatively using analytical calculations.
We then exploit numerical simulations to give a further test of the relations in 
Sec.~\ref{sec:simulations}.
We finally summarize our findings in Sec.~\ref{sec:summary}.

\section{Matter density correlations}
\label{sec:density}

In this paper, we assume that the nonlinear matter density contrast, 
$\delta(\bfx,t)=[\rho(\bfx,t)-\bar\rho]/\bar\rho$, 
is fully defined at any time by the initial linear density contrast $\delta_{L0}$
(i.e., decaying modes have had time to vanish), and that the latter is Gaussian and fully 
described by the linear power spectrum $P_{L0}(k)$,
\be
\langle \tdelta_{L0}(\bfk_1) \tdelta_{L0}(\bfk_2) \rangle = P_{L0}(k_1) \delta_D(\bfk_1+\bfk_2) ,
\label{PL0-def}
\ee
where we denote with a tilde Fourier-space fields.
The matter density contrast can also be written in terms of the
particle trajectories, $\bfx(\bfq,t)$, where $\bfq$ is the Lagrangian coordinate of the
particles, as
\be
k \neq 0 : \;\;\; \tdelta(\bfk,t) & = & \int \frac{\dd\bfx}{(2\pi)^3} \; e^{-\ii\bfk\cdot\bfx} \, \delta(\bfx,t)
\\
& = & \int \frac{\dd\bfq}{(2\pi)^3} \; e^{-\ii\bfk\cdot\bfx(\bfq,t)} ,
\label{tdelta-def}
\ee
where we discarded a Dirac term that does not contribute for $k\neq 0$.
This expression follows from the conservation of matter, $\rho \dd\bfx = \bar\rho\dd\bfq$,
which yields $[1+\delta(\bfx)] \dd\bfx = \dd\bfq$.

Using the Gaussianity of the linear density field $\delta_{L0}$,
integrations by parts allow us to write the correlation between $\ell$ linear
fields and $n$ nonlinear fields in terms of the response of the latter to changes of the
initial conditions \cite{Valageas:2014aa,Valageas:2014ab}
\be
\hspace{-0.5cm}\langle \tdelta_{L0}(\bfk'_1) .. \tdelta_{L0}(\bfk'_{\ell}) \tdelta(\bfk_1,t_1) .. \tdelta(\bfk_n,t_n) 
\rangle  & = & 
\nonumber \\
&& \hspace{-5.5cm} P_{L0}(k'_1) ..  P_{L0}(k'_{\ell}) \left \langle 
\frac{{\cal D}^{\ell}[\tdelta(\bfk_1,t_1) .. \tdelta(\bfk_n,t_n)]}
{{\cal D}\tdelta_{L0}(-\bfk'_1) .. {\cal D}\tdelta_{L0}(-\bfk'_{\ell})} \right \rangle  .
\label{Cn-Rn-1}
\ee
This exact relation, which only relies on the Gaussianity of the initial condition $\tdelta_{L0}$,
holds for any nonlinear field $\tdelta$, which is not necessarily identified with the nonlinear
density contrast.
 It is also the basis of the consistency relations between $(\ell+n)-$ and
$n-$point polyspectra, in the limit $k'_j \rightarrow 0$, when one can write the right-hand side
in terms of $\langle \tdelta(\bfk_1,t_1) .. \tdelta(\bfk_n,t_n) \rangle$ multiplied by some 
deterministic prefactors or operators
\cite{Kehagias2013,Peloso2013,Creminelli2013,Kehagias2014,Peloso2014,Creminelli2014,Valageas:2014aa,Valageas:2014ab,Kehagias2014a}.

In this paper, we extend the analysis presented in \cite{Valageas:2014ab} for the real-space
density field to the redshift-space density field.
Because of the Doppler effect associated with the peculiar velocities, the radial position
of cosmological objects (e.g., galaxies) is not exactly given by their redshift, interpreted as a 
distance within the uniform background cosmology. 
For instance, receding objects appear to have a slightly
higher redshift than the one associated with their actual location, and one is led to introduce
the redshift-space coordinate $\bfs$ defined as \cite{Kaiser:1987aa,Taylor1996,Scoccimarro2004}
\be
\bfs = \bfx + \frac{v_r}{\dot{a}} \, \bfe_r  ,
\label{s-def-1}
\ee
where $\bfe_r$ is the radial unit vector along the line of sight (we use a plane-parallel 
approximation throughout this article), $v_r$ the line-of-sight component of the peculiar 
velocity $\bfv$, and $\dot{a}=\dd a/\dd t $ the time derivative of the scale-factor $a(t)$.
Then, the redshift-space density contrast $\delta^s$ can be written in terms of the
Lagrangian coordinate $\bfq$ of the particles as 
\cite{Taylor1996,Scoccimarro2004,Valageas2011a}
\be
k \neq 0 : \;\;\; \tdelta^s(\bfk,t) = \int \frac{\dd\bfq}{(2\pi)^3} \; e^{-\ii\bfk\cdot\bfs(\bfq,t)} ,
\label{tdelta-s-def}
\ee
where we again discarded a Dirac term that does not contribute for $k\neq 0$.
This is the same expression as Eq.(\ref{tdelta-def}) for the real-space density contrast 
$\tdelta(\bfk)$, except that $\bfx(\bfq,t)$ in the exponential is replaced by $\bfs(\bfq,t)$,
and it again follows from the conservation of matter, $[1+\delta^s(\bfs)] \dd\bfs = \dd\bfq$.
Then, the redshift-space generalisation of Eq.(\ref{Cn-Rn-1})
reads as
\be
\hspace{-0.5cm} \langle \tdelta_{L0}(\bfk'_1) .. \tdelta_{L0}(\bfk'_{\ell}) \tdelta^s(\bfk_1,t_1) .. 
\tdelta^s(\bfk_n,t_n) \rangle  & = & \nonumber \\
&& \hspace{-5.5cm} P_{L0}(k'_1) .. P_{L0}(k'_{\ell})  \left \langle \frac{{\cal D}^{\ell}[\tdelta^s(\bfk_1,t_1) .. 
\tdelta^s(\bfk_n,t_n)]}{{\cal D}\tdelta_{L0}(-\bfk'_1)..{\cal D}\tdelta_{L0}(-\bfk'_{\ell})} 
\right \rangle  .
\label{Cn-Rn-s-1}
\ee

As in \cite{Valageas:2014ab,Nishimichi2014}, we focus on the relations obtained
for $\ell=1$ by performing a spherical average over the angles of the large-scale wave 
number $\bfk'$. 
This removes the leading order contribution, associated with a uniform displacement of
small-scale structures by larger-scale modes, that vanishes for equal-time statistics
($t_1=..=t_n$)
\cite{Kehagias2013,Peloso2013,Creminelli2013,Kehagias2014,Peloso2014,Creminelli2014,Valageas:2014aa}.
One is left with the next-order contribution, which does not vanish at equal times
\cite{Valageas:2014ab,Kehagias2014a,Nishimichi2014}, and is
associated with the change to the growth of small-scale structures in a perturbed mean 
density background, modulated by the larger scale modes.
In configuration space, this means that we consider angular-averaged quantities of the form
\cite{Valageas:2014ab,Nishimichi2014}
\be
C^n_W = \int \!\! \dd\bfx' W(x') \, \langle \delta_{L0}(\bfx') \delta^s(\bfs_1,t_1) .. \delta^s(\bfs_n,t_n) 
\rangle ,
\label{Cn-Wx-1}
\ee
which read in Fourier space as
\be
\tilde{C}^n_W = (2\pi)^3 \! \int \!\! \dd\bfk' \tilde{W}(k') \, \langle \tdelta_{L0}(\bfk') \tdelta^s(\bfk_1,t_1) .. 
\delta^s(\bfk_n,t_n) \rangle , \;\;
\label{Cn-Wk-1}
\ee
where $W(x')$ [and its Fourier transform $\tilde{W}(k')$] is a large-scale spherical window function.
Using Eq.(\ref{Cn-Rn-s-1}), we obtain
\be
C^n_W = \left. \frac{\dd}{\dd\varepsilon_0} \right |_{\varepsilon_0=0} 
\langle \delta^s(\bfs_1,t_1) .. \delta^s(\bfs_n,t_n) \rangle_{\varepsilon_0} ,
\label{Cns-eps-1}
\ee
and a similar relation for $\tilde{C}^n_W$, where $\langle .. \rangle_{\varepsilon_0}$ is the statistical 
average with respect to the Gaussian initial conditions $\delta_{L0}$, when the linear density field
is modified as
\be
\delta_L(\bfx) \rightarrow \delta_L(\bfx) + \varepsilon_0 D_+(t) \int \dd\bfx' W(x') 
C_{L0}(\bfx,\bfx') ,
\label{eps-x-1}
\ee
where $C_{L0}$ is the linear density contrast real-space correlation function.
In the large scale limit for the window function $W$, which corresponds to the limit
$k'\rightarrow 0$, the integral 
over $\bfx'$ is independent of the position $\bfx$ in the small-scale region, at leading 
order in the ratio of scales, and the initial linear density contrast is merely shifted
by a uniform amount,
\be
k' \rightarrow 0 : \;\;\; \Delta \delta_{L0} = \varepsilon_0 \int \dd\bfx' W(x') C_{L0}(x') .
\label{eps0-eps-1}
\ee
This corresponds to a change of the background mean density $\bar\rho$, which means
that we must obtain the impact of a small change of $\bar\rho$, hence of cosmological
parameters, on the small-scale correlation 
$\langle \delta^s(\bfs_1,t_1) .. \delta^s(\bfs_n,t_n) \rangle$.

\section{Approximate symmetry of the cosmological gravitational dynamics}
\label{sec:approx-sym}

On scales much smaller than the horizon, where the Newtonian approximation
is valid, the equations of motion read as \cite{Peebles1980}
\be
\frac{\partial\delta}{\partial t} + \frac{1}{a} \nabla \cdot [ (1+\delta) \bfv ] = 0 , 
\label{cont-1}
\ee
\be
\frac{\partial\bfv}{\partial t} + H \bfv + \frac{1}{a} (\bfv\cdot\nabla)\bfv = 
- \frac{1}{a} \nabla \phi ,
\label{Euler-1}
\ee
\be
\nabla^2 \phi = 4\pi {\cal G} \bar{\rho} a^2 \delta .
\label{Poisson-1}
\ee
Here, we use the single-stream approximation to simplify the presentation,
but our results remain valid beyond shell crossing.
Linearizing these equations over $\{\delta,\bfv\}$, one obtains the linear growth rates
$D_\pm(t)$, which are the independent solutions of 
\cite{Peebles1980,Bernardeau2002}
\be
\ddot{D} + 2 H \dot{D} - 4\pi{\cal G}\bar{\rho} D = 0 .
\label{Dlin-1}
\ee
Then, it is convenient to make the change of variables 
\cite{Crocce2006a,Crocce2006,Valageas2008,Valageas:2014ab}
\be
\eta = \ln D_+ , \;\;\; \bfv = \dot{a} f \bfu , \;\;\; \phi = (\dot{a} f)^2 \varphi ,
\label{eta-3D}
\ee
with
\be
f= \frac{\dd\ln D_+}{\dd \ln a} = \frac{a\dot{D}_+}{\dot{a}D_+} ,
\label{f-def}
\ee
and the equations of motion read as
\be
\frac{\partial\delta}{\partial\eta} + \nabla \cdot [ (1+\delta) \bfu ] = 0 , 
\label{cont-2}
\ee
\be
\frac{\partial\bfu}{\partial \eta} + \left( \frac{3\Omega_{\rm m}}{2f^2} - 1 \right) \bfu + (\bfu\cdot\nabla)\bfu 
= - \nabla \varphi ,
\label{Euler-2}
\ee
\be
\nabla^2 \varphi = \frac{3\Omega_{\rm m}}{2f^2} \, \delta .
\label{Poisson-2}
\ee
Here, $\Omega_{\rm m}(t)$ is the matter density cosmological parameter as a function of time,
which obeys $4\pi{\cal G}\bar{\rho}=(3/2) \Omega_{\rm m} H^2$.
As pointed out in \cite{Valageas:2014ab}, within the approximation $\Omega_{\rm m}/f^2 \simeq 1$
(which is used by most perturbative approaches \cite{Bernardeau2002}), 
all explicit dependence on cosmology 
disappears from the equations of motion (\ref{cont-2})-(\ref{Poisson-2}).
This means that the dependence of the density and velocity fields on cosmology is fully absorbed
by the change of variable (\ref{eta-3D}).
Then, a change of the background density, as in Eq.(\ref{eps0-eps-1}), can be absorbed
through a change of the time-dependent functions $\{a(t),D_+(t),f(t)\}$, which enter
the change of variables (\ref{eta-3D}) \cite{Valageas:2014ab}.

Here, we used the single-stream approximation to simplify the presentation,
but the results remain valid beyond shell crossing, as the dynamics of particle trajectories,
$\bfx(\bfq,t)$, follow the equation
\be
\frac{\partial^2\bfx}{\partial\eta^2} +  \left( \frac{3\Omega_{\rm m}}{2f^2} - 1 \right) 
\frac{\partial\bfx}{\partial\eta} = - \nabla \varphi ,
\label{traj-2}
\ee
where $\varphi$ is the rescaled gravitational potential (\ref{Poisson-2}).
This explicitly shows that they satisfy the same approximate symmetry.
Therefore, our results are not restricted to the perturbative regime
and also apply to small nonlinear scales governed by shell-crossing effects,
as long as the approximation $\Omega_{\rm m}/f^2 \simeq 1$ is sufficiently accurate
(but this also means that we are restricted to scales dominated by gravity).

\section{Angular-averaged consistency relations}
\label{sec:angular}

As described in \cite{Valageas:2014ab}, the impact of a small uniform change of the matter density
background can be obtained by considering two universes with nearby background densities
and scale factors, $\{\bar\rho(t),a(t)\}$ and $\{\bar\rho'(t),a'(t)\}$,
with
\be
\bar\rho a^3 = \bar\rho' a'^3 = \bar\rho_0 , \;\;
a' = a [1 \!-\! \epsilon(t) ] , \;\;
\bar\rho' = \bar\rho [ 1 \!+\! 3\epsilon(t) ] .
\label{eps-def}
\ee
Here and in the following, we only keep terms up to linear order over $\epsilon$.
Then, writing the Friedmann equations for the two scale factors $a(t)$ and $a'(t)$
and linearizing over $\epsilon$, we find that $\epsilon(t)$ must satisfy the same
equation (\ref{Dlin-1}) as the linear growing mode \cite{Peebles1980}. Thus, we can write
\be
\epsilon(t) =  D_+(t) \, \epsilon_0 .
\label{eps0-def}
\ee

For our purposes, the universe $\{\bar\rho(t),a(t)\}$ is the actual universe, with the zero-mean
initial condition $\delta_{L0}$, to which is added the uniform density perturbation 
(\ref{eps0-eps-1}). To recover zero-mean density fluctuations, we must shift the background
by the same amount. Thus, this new background $\{\bar\rho'(t),a'(t)\}$ is given by
\be
\epsilon_0  =  \frac{1}{3} \Delta \delta_{L0} = \frac{\varepsilon_0}{3} \int \dd\bfx' W(x') C_{L0}(x') ,
\label{eps-vareps0}
\ee
where we used the last relation (\ref{eps-def}), which gives at linear order over $\delta$ and
$\epsilon$, $\delta_L= \delta_L'+3\epsilon$.

Because both frames refer to the same physical system, we have
$\bfr' = \bfr = a' \bfx' = a \bfx$, $\bar\rho' (1+\delta') = \bar\rho (1+\delta)$,
where $\bfr=\bfr'$ is the physical coordinate. Thus, we have the relations
\be
\bfx' = (1+\epsilon) \bfx , \;\;\; 
\delta' = \delta - 3 \epsilon (1+\delta) , \;\;\;
\bfv' = \bfv + \dot{\epsilon} a \bfx , 
\label{x-d-v-1}
\ee
where we used Eq.(\ref{eps-def}) and only kept terms up to linear order over 
$\epsilon$.
In particular, we can check that if the fields $\{\delta',\bfv',\phi'\}$ satisfy the equations of 
motion (\ref{cont-1})-(\ref{Poisson-1}) in the primed frame, the fields 
$\{\delta,\bfv,\phi\}$ satisfy the equations of motion (\ref{cont-1})-(\ref{Poisson-1})
in the unprimed frame, with the gravitational potential transforming as
$\phi' = \phi - a^2 (\ddot{\epsilon} + 2 H \dot{\epsilon} ) x^2/2$.
This remains valid beyond shell crossing: if the trajectories $\bfx'(\bfq,t)$ satisfy the
equation of motion in the primed frame, the trajectories $\bfx(\bfq,t)=
(1-\epsilon)\bfx'(\bfq,t)$ satisfy the equation of motion in the unprimed frame.

From the definition (\ref{s-def-1}) and Eq.(\ref{x-d-v-1}), we obtain the relation between the 
redshift-space coordinates,
\be
\bfs' = (1+\epsilon) \bfs + \frac{\dot{\epsilon}}{H} s_r \, \bfe_r ,
\label{sp-s}
\ee
using $a'=(1-\epsilon) a$ and $H'=H-\dot{\epsilon}$.
Then, using for instance the expressions (\ref{tdelta-def}) and (\ref{tdelta-s-def}), 
the real-space and the redshift-space density contrasts in the actual unprimed frame, 
with the uniform overdensity $\Delta\delta_{L0}=3\epsilon_0$, can be written as 
\cite{Valageas:2014ab}
\be
k \neq 0 : \;\;\; \tdelta_{\epsilon_0}(\bfk,t) = \tdelta[(1-\epsilon) \bfk , D_{+\epsilon_0} ] ,
\label{map-k-1}
\ee
and
\be
\tdelta^s_{\epsilon_0}(\bfk,t) = \tdelta^s[(1-\epsilon) \bfk - \frac{\dot{\epsilon}}{H} k_r \, \bfe_r , 
D_{+\epsilon_0} , f_{\epsilon_0}] ,
\label{map-s-k-1}
\ee
where we disregarded a Dirac factor $\delta_D(\bfk)$ that does not contribute for $k \neq 0$.
In Eqs.(\ref{map-k-1})-(\ref{map-s-k-1}), the subscript ``$\epsilon_0$'' recalls that we consider
the formation of large-scale structures in the actual universe to which is added the small uniform
overdensity $\Delta\delta_{L0}=3\epsilon_0$.

The physical meaning of the expression (\ref{map-k-1}) directly follows from the mapping
(\ref{x-d-v-1}) and the independence on cosmology of the equations of motion
(\ref{cont-2})-(\ref{Poisson-2}), within the approximation $\Omega_{\rm m}/f^2 \simeq 1$.
It means that in the primed universe, with the slightly higher background density
$\bar\rho' = (1+3\epsilon) \bar\rho$ (focusing for instance on the case $\epsilon>0$), 
comoving distances $\bfx'$ show the small isotropic dilatation (\ref{x-d-v-1})
[because the higher background density yields a higher gravitational force and a smaller 
scale factor $a'(t)$],
whence an isotropic contraction of wave numbers $\bfk'$, while the linear growth factor $D_+(t)$
is also modified. Moreover, the approximate symmetry discussed in Sec.~\ref{sec:approx-sym}
implies that all time and cosmological dependence can be absorbed through the time coordinate
$D_+$, if we work with the rescaled field $\{\delta,\bfu,\varphi\}$ of Eq.(\ref{eta-3D}).
This is denoted by the rescaled time coordinate $D_{+\epsilon_0}$ in the right-hand side of
Eq.(\ref{map-k-1}), where the subscript ``$\epsilon_0$'' recalls that we must take into account
the impact of the modified background onto the linear growth factor.

For the redshift-space density contrast (\ref{map-s-k-1}) two new effects arise, as compared
with the real-space density contrast (\ref{map-k-1}).
First, the mapping $\bfs \leftrightarrow \bfs'$ is no longer isotropic because of the
peculiar velocity component along the line of sight, see Eq.(\ref{sp-s}), which also leads to
an anisotropic relationship $\bfk \leftrightarrow \bfk'$.
Second, in addition to the time coordinate $D_+$, the redshift-space density contrast involves
the new quantity $f(t)$.
This follows from the definition (\ref{s-def-1}),
which can be written in terms of the rescaled velocity field $\bfu$ of Eq.(\ref{eta-3D}) as
\be
\bfs = \bfx + \frac{v_r}{\dot{a}} \, \bfe_r = \bfx + f u_r \, \bfe_r .
\label{s-def}
\ee
This shows that in addition to the rescaled term $u_r$, which only depends on time and
cosmology through the time coordinate $D_+$, within the approximate symmetry of 
Sec.~\ref{sec:approx-sym}, the line-of-sight component explicitly involves a 
time and cosmology dependent factor $f(t)$, which must be taken into account 
in Eq.(\ref{map-s-k-1}).

Then, to derive the angular-averaged consistency relations through 
Eq.(\ref{Cns-eps-1}), we simply need to use Eq.(\ref{map-s-k-1}) to obtain the
derivative of the redshift-space density contrast with respect to $\epsilon_0$,
and next to use Eq.(\ref{eps-vareps0}).
This yields
\be
\frac{\partial \tdelta^s(\bfk)}{\partial\epsilon_0} & = & 
\frac{\partial D_{+\epsilon_0}}{\partial\epsilon_0} \frac{\partial \tdelta^s}{\partial D_+} 
+ \frac{\partial f_{\epsilon_0}}{\partial\epsilon_0} \frac{\partial \tdelta^s}{\partial f} \nonumber \\
&& - D_+(t) \bfk \cdot \frac{\partial\tdelta^s}{\partial\bfk}
- f D_+ k_r \frac{\partial\tdelta^s}{\partial k_r} ,
\label{d-delta-d-eps0-1}
\ee
where we disregarded a Dirac factor that does not contribute for wave numbers
$\bfk\neq 0$.

As found in \cite{Baldauf2011,Valageas:2014ab}, the derivative of the linear growth factor 
reads as
\be
\left. \frac{\partial D_{+\epsilon_0}}{\partial\epsilon_0} \right |_{\epsilon_0=0} = \frac{13}{7} D_+(t)^2 .
\label{d-D+epsilon0}
\ee
This corresponds to $D_+'= D_+ + (13/7) D_+^2 \epsilon_0$ for the linear growing mode in the
primed frame, while $a'= a -D_+ a \epsilon_0$ and $H'=H-\dot{D}_+\epsilon_0$.
From the definition (\ref{f-def}), we obtain $f' = f + f (13/7 D_+ + \dot{D}_+/H) \epsilon_0$,
whence
\be
\left. \frac{\partial f_{\epsilon_0}}{\partial\epsilon_0} \right |_{\epsilon_0=0} = 
f D_+ \left( \frac{13}{7} + f \right) .
\label{d-fepsilon0}
\ee
Therefore, Eq.(\ref{d-delta-d-eps0-1}) gives
\be
\frac{\partial \tdelta^s(\bfk)}{\partial\epsilon_0} & = & 
\frac{13}{7} D_+(t)^2 \frac{\partial \tdelta^s}{\partial D_+} 
+ f D_+ \left( \frac{13}{7} + f \right) \frac{\partial \tdelta^s}{\partial f} \nonumber \\
&& - D_+(t) \bfk \cdot \frac{\partial\tdelta^s}{\partial\bfk}
- f D_+ k_r \frac{\partial\tdelta^s}{\partial k_r} .
\label{d-deltak-d-eps0-2}
\ee
Of course, when we set $f$ to zero, we recover the expression of the derivative with respect
to $\epsilon_0$ of the real-space density contrast $\tdelta$ \cite{Valageas:2014ab}.
In configuration space, this reads as
\be
\frac{\partial \delta^s(\bfs)}{\partial\epsilon_0} & = & D_+ \left[ 
\frac{13}{7} \frac{\partial}{\partial \ln D_+} 
+ \left( \frac{13}{7} + f \right) f \frac{\partial}{\partial f} \right. \nonumber \\
&& \hspace{-0.cm} \left. + 3+f  + \bfs \cdot \frac{\partial}{\partial\bfs}
+ f s_r \frac{\partial}{\partial s_r} \right] \delta^s(\bfs) .
\label{d-delta-d-eps0-2}
\ee

Next, from Eqs.(\ref{Cns-eps-1}) and (\ref{eps-vareps0}), we obtain as for the real-space
correlations \cite{Valageas:2014ab},
\be
C_W^n & \!\! = \!\! & \int \dd\bfx' W(x') C_{L0}(x') \sum_{i=1}^n \frac{D_{+i}}{3} 
\biggl [ 3 + f_i + \frac{13}{7}  \frac{\partial}{\partial \ln D_{+i}} \;\;\;  \nonumber \\
&& \hspace{-0.5cm} + \left( \frac{13}{7} + f_i \right) f_i \frac{\partial}{\partial f_i}
+ \biggl ( \bfs_i - \frac{1}{n} \sum_{j=1}^n \bfs_j \biggl ) \cdot 
\frac{\partial}{\partial\bfs_i} \nonumber \\
&& \hspace{-0.5cm} + f_i \biggl ( s_{ri} - \frac{1}{n} \sum_{j=1}^n s_{rj} \biggl )
\frac{\partial}{\partial s_{ri}} \biggl ] \langle \delta^s(\bfs_1,t_1) .. \delta(\bfs_n,t_n) \rangle .
\label{Cn-x-eps0-3}
\ee
The counter terms of the form $-1/n \sum_j \bfs_j$ ensure that all expressions are invariant
with respect to uniform translations [by explicitly setting the small-scale region at the
center of the large-scale perturbation (\ref{eps-x-1})].
They are irrelevant for equal-time statistics, $t_1=..=t_n$, where factors of the form
$\sum_i\bfs_i\cdot\frac{\partial}{\partial\bfs_i} \langle \tdelta^s_1 .. \tdelta^s_n\rangle$
are already invariant with respect to uniform translations.

The comparison with Eq.(\ref{Cn-Wx-1}) gives, after writing the correlations in terms of
Fourier-space polyspectra,
\be
\int \frac{\dd \Omega_{\bfk'}}{4\pi} \langle \tdelta_{L0}(\bfk') \tdelta^s(\bfk_1,t_1) .. 
\tdelta^s(\bfk_n,t_n) \rangle'_{k'\rightarrow 0} =  P_{L0}(k') \nonumber \\
&& \hspace{-7.9cm} \times \sum_{i=1}^n D_{+i} \biggl [ \frac{1}{n} + \frac{f_i}{3n} 
+ \frac{13}{21} \frac{\partial}{\partial\ln D_{+i}} + \left( \frac{13}{7} + f_i \right) \frac{f_i}{3}
 \frac{\partial}{\partial f_i} \nonumber \\
&& \hspace{-7.9cm}  - \sum_{j=1}^n (\delta^K_{i,j} - \frac{1}{n} ) 
\frac{\bfk_i}{3} \! \cdot \! \frac{\partial}{\partial\bfk_j} 
- f_i \sum_{j=1}^n (\delta^K_{i,j} - \frac{1}{n} ) 
\frac{k_{ri}}{3} \frac{\partial}{\partial k_{rj}} \biggl ]
\nonumber \\
&& \hspace{-7.9cm} \times \; \langle \tdelta^s(\bfk_1,t_1) .. \tdelta^s(\bfk_n,t_n) \rangle' ,
\label{tCn0-1}
\ee
where $\Omega_{\bfk'}$ is the unit vector along the direction of $\bfk'$ and
$\delta^K_{i,j}$ the Kronecker symbol.
The subscript $k'\rightarrow 0$ recalls that this relation only gives the leading-order term
in the large-scale limit $k' \rightarrow 0$, whereas the wave numbers $\{\bfk_1,..,\bfk_n\}$
are fixed and may be within the nonlinear regime.
Here we denoted with a prime the reduced polyspectra, defined as
\be
\langle \tdelta^s(\bfk_1) .. \tdelta^s(\bfk_n) \rangle =
\langle \tdelta^s(\bfk_1) .. \tdelta^s(\bfk_n) \rangle'  \; \delta_D(\bfk_1 \!+\! .. \!+\! \bfk_n) , \;\;\;
\label{multi-spectra}
\ee
where we explicitly factor out the Dirac factor associated with statistical homogeneity.
In particular, this means that $\langle \tdelta^s(\bfk_1) .. \tdelta^s(\bfk_n) \rangle'$ can be
written as a function of the $n-1$ wave numbers $\{\bfk_1,..,\bfk_{n-1}\}$ only.

On large scales we recover the linear theory \cite{Kaiser:1987aa,Bernardeau2002}, with 
$\tdelta(\bfk',t') \simeq D_+(t') \tdelta_{L0}(\bfk')$ and 
$\tdelta^s(\bfk',t') \simeq D_+(t') \tdelta_{L0}(\bfk') (1+f'\mu'^2)$,
where $\mu'$ is the cosine of the wave number $\bfk'$ with the line of sight, as in
\be
\mu= \frac{\bfk \cdot \bfe_r}{k} .
\label{mu-def}
\ee
Therefore, Eq.(\ref{tCn0-1}) also gives
\be
\int \! \frac{\dd \Omega_{\bfk'}}{4\pi} \biggl \langle \frac{\tdelta^s(\bfk',t')}{1+f'\mu'^2} 
\tdelta^s(\bfk_1,t_1) .. \tdelta^s(\bfk_n,t_n) \! \biggl \rangle'_{k'\rightarrow 0} =  P_L(k',t') 
\nonumber \\
&& \hspace{-8.8cm} \times \sum_{i=1}^n \frac{D_{+i}}{D_+'} \biggl [ \frac{1}{n} + \frac{f_i}{3n} 
+ \frac{13}{21} \frac{\partial}{\partial\ln D_{+i}} + \left( \frac{13}{7} + f_i \right) \frac{f_i}{3}
 \frac{\partial}{\partial f_i} \nonumber \\
&& \hspace{-8.8cm}  - \sum_{j=1}^n (\delta^K_{i,j} - \frac{1}{n} ) 
\frac{\bfk_i}{3} \! \cdot \! \frac{\partial}{\partial\bfk_j} 
- f_i \sum_{j=1}^n (\delta^K_{i,j} - \frac{1}{n} ) 
\frac{k_{ri}}{3} \frac{\partial}{\partial k_{rj}} \biggl ]
\nonumber \\
&& \hspace{-8.8cm} \times \; \langle \tdelta^s(\bfk_1,t_1) .. \tdelta^s(\bfk_n,t_n) \rangle' .
\label{tCn0-2}
\ee

When all times are equal, $t'=t_1=..=t_n\equiv t$, this simplifies as
\be
\hspace{-0.5cm} \int \! \frac{\dd \Omega_{\bfk'}}{4\pi} \biggl \langle \frac{\tdelta^s(\bfk')}{1+f\mu'^2} 
\tdelta^s(\bfk_1) .. \tdelta^s(\bfk_n) \! \biggl \rangle'_{k'\rightarrow 0} =  P_L(k')  \nonumber \\
&& \hspace{-7.cm} \times \biggl [ 1+ \frac{f}{3} + \frac{13}{21} \frac{\partial}{\partial\ln D_+} 
+ \left( \frac{13}{7} \!+\! f \right) \frac{f}{3}
 \frac{\partial}{\partial f}  \nonumber \\
&& \hspace{-7.cm} - \sum_{i=1}^n \frac{k_i}{3} \frac{\partial}{\partial k_i} 
 - f \sum_{i=1}^n \frac{k_{ri}}{3} \frac{\partial}{\partial k_{ri}} \biggl ]
\langle \tdelta^s(\bfk_1) .. \tdelta^s(\bfk_n) \rangle' . \;\; 
\label{tCn0-3}
\ee

\section{Bispectrum}
\label{sec:bspectrum}

\subsection{Relation in $\{k,\mu^2\}$ space}
\label{sec:k-mu2}

The lowest-order equal-time consistency relation obtained from Eq.(\ref{tCn0-3}) 
corresponds to $n=2$, that is, the bispectrum built from the correlation between two
small-scale modes and one large-scale mode.
We define the bispectrum as in Eq.(\ref{multi-spectra}),
\be
\langle \tdelta^s(\bfk_1) \tdelta^s(\bfk_1) \tdelta^s(\bfk_1) \rangle & = & B^s(\bfk_1,\bfk_2,\bfk_3)
\nonumber \\
&& \times \delta_D( \bfk_1 + \bfk_2 + \bfk_3) .
\label{bispectrum-def}
\ee
In contrast with the real-space bispectrum, $B(k_1,k_2,k_3)$, which only depends on the
lengths of the three wave numbers $\{\bfk_1,\bfk_2,\bfk_3\}$ thanks to statistical
isotropy, the redshift-space bispectrum also depends on angles because the velocity
component along the line of sight breaks the isotropy.
Then, Eq.(\ref{tCn0-3}) yields
\be
\hspace{-0.5cm} \int \! \frac{\dd \Omega_{\bfk'}}{4\pi} \frac{B^s(\bfk',\bfk-\bfk'/2,-\bfk-\bfk'/2)_{k'\rightarrow 0}}
{1+f\mu'^2} & = & P_L(k') \nonumber \\
&& \hspace{-6.4cm} \times \left[1+ \frac{f}{3} + \frac{13}{21}\frac{\partial}{\partial\ln D_+} 
+ \left( \frac{13}{7} + f \right) \frac{f}{3} \frac{\partial}{\partial f} \right.
\nonumber \\
&& \hspace{-6.4cm} \left. - \frac{1+f\mu^2}{3}\frac{\partial}{\partial\ln k}
- \frac{f}{3} 2\mu^2 (1-\mu^2) \frac{\partial}{\partial \mu^2} \right] P^s(k,\mu^2) . \;\;
\label{eq:bispec}
\ee
Here we used the symmetries of the redshift-space power spectrum to write $P^s(\bfk)$ 
as a function of $k$ and $\mu^2$.
In Eq.(\ref{eq:bispec}), the power spectrum is written as a function of time through the functions
$D_+$ and $f$, that is, 
\be
P^s(\bfk;t) = P^s(k,\mu^2;D_+,f) .
\ee
In particular, in the linear regime we have the well-known expression
\be
P^s_L(\bfk;t) = D_+^2 P_{L0}(k) (1+f \mu^2)^2 ,
\label{PL-def}
\ee
where $P_{L0}$ is the linear real-space power spectrum today.
When $f=0$, the relation (\ref{eq:bispec}) recovers the real-space consistency relation,
as it should.

\subsection{Multipole expansion}
\label{sec:multipole}

The consistency relation (\ref{eq:bispec}) is written for a given value of $k$ and $\mu$.
In practice, rather than considering the redshift-space power spectrum over a grid 
of $\mu$, one often expands the dependence on $\mu$ over Legendre polynomials.
Thus, we write the nonlinear redshift-space power spectrum as
\be
P^s(\bfk) = P^s(k,\mu^2) = \sum_{\ell=0}^{\infty} P^s_{2\ell}(k) \, L_{2\ell}(\mu) ,
\label{Ps-Leg}
\ee
where $L_{\ell}(\mu)$ is the Legendre polynomial of order $\ell$. Only even orders contribute
to this expansion because $P^s$ is an even function of $\mu$.
Substituting into Eq.(\ref{eq:bispec}), we obtain
\be
\hspace{-0.cm} \frac{4\ell \!+\! 1}{2} \int_{-1}^{1} \!\! \dd\mu \, L_{2\ell}(\mu) 
\int \! \frac{\dd \Omega_{\bfk'}}{4\pi} 
\frac{B^s(\bfk',\bfk \!-\! \frac{\bfk'}{2},-\bfk \!-\! \frac{\bfk'}{2})_{k'\rightarrow 0}}
{1+f\mu'^2} && \nonumber \\
&& \hspace{-8.3cm} = P_L(k') \biggl \lbrace \left[1+ \frac{f}{3} 
+ \frac{13}{21}\frac{\partial}{\partial\ln D_+} 
+ \left( \frac{13}{7} + f \right) \frac{f}{3} \frac{\partial}{\partial f} \right.
\nonumber \\
&& \hspace{-8cm} \left. - \frac{1}{3}\frac{\partial}{\partial\ln k} \right] P^s_{2\ell}(k) 
- \frac{f}{3} \frac{\partial}{\partial\ln k} \left[ \frac{(2\ell \!-\! 1)2\ell}{(4\ell \!-\! 3)(4\ell \!-\! 1)}
P^s_{2\ell-2} \right. \nonumber \\
&& \hspace{-8cm} \left. +  \frac{(8\ell^2 \!+\! 4\ell \!-\! 1)}{(4\ell \!-\! 1)(4\ell \!+\! 3)} P^s_{2\ell}
+ \frac{(2\ell \!+\! 1)(2\ell \!+\! 2)}{(4\ell \!+\! 3)(4\ell \!+\! 5)} P^s_{2\ell+2} \right] \nonumber \\
&& \hspace{-8cm} - \frac{f}{3} \left[ - \frac{(2\ell \!-\! 2) (2\ell \!-\! 1) 2\ell}{(4\ell \!-\! 3)(4\ell \!-\! 1)}
P^s_{2\ell-2} +  \frac{(2\ell (2\ell \!+\! 1)}{(4\ell \!-\! 1)(4\ell \!+\! 3)} P^s_{2\ell} \right. \nonumber \\
&& \hspace{-8cm} \left. + \frac{(2\ell \!+\! 1)(2\ell \!+\! 2)(2\ell \!+\! 3)}{(4\ell \!+\! 3)(4\ell \!+\! 5)} 
P^s_{2\ell+2} \right] \biggl \rbrace .
\label{eq:bispec-ell}
\ee
For the first two multipoles, $2\ell=0$ and $2\ell=2$, this yields
\be
\hspace{-0.cm} \int_{-1}^{1} \!\! \frac{\dd\mu}{2} \int \! \frac{\dd \Omega_{\bfk'}}{4\pi} 
\frac{B^s_{k'\rightarrow 0}}{1+f\mu'^2} = P_L(k') \biggl \lbrace 
\left[1+ \frac{f}{3} + \frac{13}{21}\frac{\partial}{\partial\ln D_+} \right. && \nonumber \\
&& \hspace{-7.9cm}  \left. + \left( \frac{13}{7} \!+\! f \right) \frac{f}{3} \frac{\partial}{\partial f} 
- \frac{1}{3}\frac{\partial}{\partial\ln k} \right] P^s_0(k)
- \frac{2f}{15} P^s_2(k) \nonumber \\
&& \hspace{-7.9cm} - \frac{f}{3} \frac{\partial}{\partial\ln k} \left[ \frac{1}{3} P^s_0(k)
+  \frac{2}{15} P^s_2(k) \right]  \biggl \rbrace ,
\label{eq:bispec-ell-0}
\ee
and
\be
\hspace{-0.cm} \frac{5}{2} \int_{-1}^{1} \!\! \dd\mu \, L_2(\mu) \int \! \frac{\dd \Omega_{\bfk'}}{4\pi} 
\frac{B^s_{k'\rightarrow 0}}{1+f\mu'^2} = P_L(k') \biggl \lbrace 
\left[1+ \frac{f}{3}  \right. && \nonumber \\
&& \hspace{-7.cm}  \left. + \frac{13}{21}\frac{\partial}{\partial\ln D_+}
+ \left( \frac{13}{7} \!+\! f \right) \frac{f}{3} \frac{\partial}{\partial f} 
- \frac{1}{3}\frac{\partial}{\partial\ln k} \right] P^s_2(k) \nonumber \\
&& \hspace{-7.cm} - \frac{f}{3} \frac{\partial}{\partial\ln k} \left[ \frac{2}{3} P^s_0(k)
+  \frac{11}{21} P^s_2(k) + \frac{4}{21} P^s_4(k) \right] \nonumber \\
&& \hspace{-7.cm} - \frac{f}{3} \left[ \frac{2}{7} P^s_2(k) 
+  \frac{20}{21} P^s_4(k) \right]  \biggl \rbrace .
\label{eq:bispec-ell-2}
\ee

\subsection{f-derivative}
\label{sec:f-derivative}

\subsubsection{Relations in $\{k,\mu^2\}$ space}
\label{sec:k-mu2-f}

In practice, we cannot directly measure the derivative with respect to $f$ of the redshift-space
power spectrum, because the time derivative combines the derivatives with respect to $D_+$ and
$f$.
Therefore, the expression (\ref{eq:bispec}) can only be applied to analytical models, where
the dependences on $D_+$ and $f$ are explicitly known.
To obtain an expression that can be applied to numerical or observed power spectra, we must
write the derivative with respect to $f$ in terms of observed time or space coordinates.
Since the redshift-space power spectrum must coincide with the real-space power spectrum
when either $f$ or $\mu^2$ vanishes, each factor $f$ (resp. $\mu^2$) must appear in combination
with a power of $\mu^2$ (resp. $f$).
Here we make the ansatz that the dependence on $f$ and $\mu^2$ only appears through the
combination $f\mu^2$, which is exact at the linear order (\ref{PL-def}) (but at higher orders
terms of the form $f\mu^4$, $f\mu^6$, ..., might appear).
This gives
\be
P^s(\bfk,t) = P^s(k,f\mu^2;D_+) \;\; \mbox{implies} \;\;
f \frac{\partial P^s}{\partial f} = \mu^2 \frac{\partial P^s}{\partial \mu^2} .
\label{df-dmu2-def}
\ee
This allows us to write Eq.(\ref{eq:bispec}) as
\be
\int \frac{\dd\Omega_{\bfk'}}{4\pi} \frac{B^s(\bfk',\bfk-\bfk'/2,-\bfk-\bfk'/2)}{1+f\mu'^2} & = & P_L(k')
\nonumber \\
&& \hspace{-6cm} \times \left[1+ \frac{f}{3} + \frac{13}{21}\frac{\partial}{\partial\ln D_+} 
- \frac{1+f\mu^2}{3}\frac{\partial}{\partial\ln k} \right. \nonumber \\
&& \hspace{-6cm} \left. + \left( \frac{13}{7} - f 
+ 2 f \mu^2 \right) \frac{\mu^2}{3} \frac{\partial}{\partial \mu^2} \right] P^s(k,f\mu^2;D_+) .
\label{eq:bispec-no-f}
\ee

In practice, we only measure the dependence of the power spectrum with respect to time $t$,
or scale factor $a(t)$, and wave number coordinates $\{k,\mu\}$.
Then, writing $\partial/\partial t = \dot{D}_+ \partial/\partial D_+ + \dot{f} \partial/\partial f$,
and using Eq.(\ref{df-dmu2-def}), we obtain
\be
\frac{\partial P^s}{\partial\ln D_+} = \frac{1}{f} \left[ \frac{\partial P^s}{\partial\ln a} 
- \frac{a \dot{f}}{\dot{a}f} \mu^2 \frac{\partial P^s}{\partial\mu^2} \right] ,
\label{D+_a}
\ee
which gives
\be
\hspace{-1cm} \int \frac{\dd\Omega_{\bfk'}}{4\pi} \frac{B^s(\bfk',\bfk-\bfk'/2,-\bfk-\bfk'/2)}{1+f\mu'^2} 
& = & P_L(k') \nonumber \\
&& \hspace{-6cm} \times \left[1+ \frac{f}{3} + \frac{13}{21 f}\frac{\partial}{\partial\ln a} 
- \frac{1+f\mu^2}{3}\frac{\partial}{\partial\ln k} \right. \nonumber \\
&& \hspace{-6cm} \left. + \left( \frac{13}{7} -f + 2 f \mu^2 
- \frac{13 a \dot{f}}{7 \dot{a} f^2} \right) \frac{\mu^2}{3} \frac{\partial}{\partial \mu^2} \right] 
P^s(k,\mu^2;a) .
\label{eq:bispec-a}
\ee

Using the approximation $\Omega_{\rm m}/f^2 \simeq 1$, we might simplify
Eq.(\ref{eq:bispec-a}) by writing $\dot{f} \simeq \frac{\dot{D}_+}{D_+} [- 2 + f/2 + 3f^2/2 ]$.
However, this introduces an additional source of error, and at redshift $z=0.35$, this gives
a $15\%$ error on $\dot{f}$.
We checked numerically that this can lead to violations of the consistency relations by factors
as large as $3$ or as small as $0.5$. Therefore, we keep the expression 
(\ref{eq:bispec-a}) in the following.
[The impact of the approximation $\Omega_{\rm m}/f^2 \simeq 1$ is greater on the explicit
factor $\dot{f}$ in Eq.(\ref{eq:bispec-a}) than on the consistency relation itself, which also
relied on this approximation, because the factor $\dot{f}$ is evaluated at the observed
redshift whereas the consistency relation involves the behavior of the growing modes over
all previous redshifts, following the growth of density fluctuations, which damps the impact
of late-time behaviors.]

\subsubsection{Multipole expansions}
\label{sec:multipole-f}

We can again write the relations (\ref{eq:bispec-no-f}) and (\ref{eq:bispec-a})
in terms of the multipole expansion (\ref{Ps-Leg}).
For the first two multipoles, Eq.(\ref{eq:bispec-no-f}) leads to
\be
\hspace{-0.1cm} \int_{-1}^{1} \!\! \frac{\dd\mu}{2} \int \! \frac{\dd \Omega_{\bfk'}}{4\pi} 
\frac{B^s_{k'\rightarrow 0}}{1+f\mu'^2} = P_L(k') \biggl \lbrace 
\left[1+ \frac{f}{3} + \frac{13}{21}\frac{\partial}{\partial\ln D_+} \right. && \nonumber \\
&& \hspace{-8.2cm}  \left. - \frac{1}{3}\frac{\partial}{\partial\ln k} \right] P^s_0(k)
- \frac{f}{3} \frac{\partial}{\partial\ln k} \left[ \frac{1}{3} P^s_0(k) +  \frac{2}{15} P^s_2(k) \right]  
\nonumber \\
&& \hspace{-8.2cm} + \frac{65+7f}{210} P^s_2(k) 
+ \frac{13+7f}{42} \sum_{\ell=2}^{\infty} P^s_{2\ell}(k) \biggl \rbrace ,
\label{eq:bispec-ell-0-f1}
\ee
and
\be
\hspace{-0.cm} \frac{5}{2} \int_{-1}^{1} \!\! \dd\mu \, L_2(\mu) \int \! \frac{\dd \Omega_{\bfk'}}{4\pi} 
\frac{B^s_{k'\rightarrow 0}}{1+f\mu'^2} = P_L(k') \biggl \lbrace 
\left[1+ \frac{f}{3}  \right.  \;\;\;\;\;\; && \nonumber \\
&& \hspace{-8.cm}  \left. + \frac{13}{21}\frac{\partial}{\partial\ln D_+}
- \frac{1}{3}\frac{\partial}{\partial\ln k} \right] P^s_2(k) - \frac{f}{3} \frac{\partial}{\partial\ln k}
\times \nonumber \\
&& \hspace{-8.cm} \left[ \frac{2}{3} P^s_0(k)
+  \frac{11}{21} P^s_2(k) + \frac{4}{21} P^s_4(k) \right] 
+ \frac{13+5f}{21} P^s_2(k)
\nonumber \\
&& \hspace{-8.cm} + \frac{195+65f}{126} P^s_4(k) 
+ \frac{65+35f}{42} \sum_{\ell=3}^{\infty} P^s_{2\ell}(k)
  \biggl \rbrace ,
\label{eq:bispec-ell-2-f1}
\ee
while Eq.(\ref{eq:bispec-a}) leads to
\be
\hspace{-0.cm} \int_{-1}^{1} \!\! \frac{\dd\mu}{2} \int \! \frac{\dd \Omega_{\bfk'}}{4\pi} 
\frac{B^s_{k'\rightarrow 0}}{1+f\mu'^2} = P_L(k') \biggl \lbrace 
\left[1+ \frac{f}{3} + \frac{13}{21 f}\frac{\partial}{\partial\ln a} \right. && \nonumber \\
&& \hspace{-8.2cm}  \left. - \frac{1}{3}\frac{\partial}{\partial\ln k} \right] P^s_0(k)
- \frac{f}{3} \frac{\partial}{\partial\ln k} \left[ \frac{1}{3} P^s_0(k) +  \frac{2}{15} P^s_2(k) \right] 
\nonumber \\
&& \hspace{-8.2cm} + \left( \frac{13}{42} \!+\! \frac{f}{30} \!-\! 
\frac{13 a \dot{f}}{42 \dot{a} f^2} \right) P^s_2(k) + \left( \frac{13}{42} \!+\! \frac{f}{6} \!-\! 
\frac{13 a \dot{f}}{42 \dot{a} f^2} \right) \nonumber \\
&& \hspace{-8.2cm} \times \sum_{\ell=2}^{\infty} P^s_{2\ell}(k) \biggl \rbrace ,
\label{eq:bispec-ell-0-f2}
\ee
and
\be
\hspace{-0.1cm} \frac{5}{2} \int_{-1}^{1} \!\! \dd\mu \, L_2(\mu) \int \! \frac{\dd \Omega_{\bfk'}}{4\pi} 
\frac{B^s_{k'\rightarrow 0}}{1+f\mu'^2} = P_L(k') \biggl \lbrace 
\left[1+ \frac{f}{3}  \right.  \;\;\; && \nonumber \\
&& \hspace{-7.8cm}  \left. + \frac{13}{21 f}\frac{\partial}{\partial\ln a}
- \frac{1}{3}\frac{\partial}{\partial\ln k} \right] P^s_2(k) - \frac{f}{3} \frac{\partial}{\partial\ln k}
\times \nonumber \\
&& \hspace{-7.8cm} \left[ \frac{2}{3} P^s_0(k)
+  \frac{11}{21} P^s_2(k) + \frac{4}{21} P^s_4(k) \right] + \nonumber \\
&& \hspace{-7.8cm} \left( \frac{13}{21} \!+\! \frac{5 f}{21} \!-\! 
\frac{13 a \dot{f}}{21 \dot{a} f^2} \right) P^s_2(k)
+  \left( \frac{65}{42} \!+\! \frac{65 f}{126} \!-\! \frac{65 a \dot{f}}{42 \dot{a} f^2} \right) 
\nonumber \\
&& \hspace{-7.8cm} \times P^s_4(k) + \left( \frac{65}{42} \!+\! \frac{5 f}{6} \!-\! 
\frac{65 a \dot{f}}{42 \dot{a} f^2} \right) \sum_{\ell=3}^{\infty} 
P^s_{2\ell}(k) \biggl \rbrace .
\label{eq:bispec-ell-2-f2}
\ee
As compared with Eqs.(\ref{eq:bispec-ell-0}) and (\ref{eq:bispec-ell-2}),
these relations involve all multipoles $P^s_{2\ell}$ in the right hand sides,
because the substitution (\ref{df-dmu2-def}) gives rise to factors $\mu^2 \partial/\partial\mu^2$
rather than the factor $(1-\mu^2) \partial/\partial\mu^2$ that appeared in
Eq.(\ref{eq:bispec}).
In practice, it is not possible to measure or compute all multipoles and one must truncate
these multipole series at some order $\ell_{\rm max}$.
This implies an additional approximation onto these relations 
(\ref{eq:bispec-ell-0-f1})-(\ref{eq:bispec-ell-2-f2}).

\section{Explicit checks}
\label{sec:checks}

The angular-averaged consistency relations (\ref{tCn0-1})-(\ref{tCn0-3}) 
are valid at all orders of perturbation theory and also beyond the
perturbative regime, including shell-crossing effects, within the accuracy
of the approximation $\Omega_{\rm m}/f^2 \simeq 1$ (and as long as gravity is the
dominant process).

We now provide two explicit checks of the angular-averaged consistency relations 
(\ref{tCn0-1})-(\ref{tCn0-3}).
First, we check these relations for the lowest-order case $n=2$, that is, for the
bispectrum, at lowest order of perturbation theory.
Second, we present a fully nonlinear and nonperturbative check, for arbitrary
$n-$point polyspectra, in the one-dimensional case.

\subsection{Perturbative check}
\label{sec:perturbative}

Here we briefly check the consistency relations for the
lowest order case, $n=2$, given by Eq.(\ref{eq:bispec}) at equal times, at lowest order of
perturbation theory.
At this order, the equal-time redshift-space matter density bispectrum reads as 
\cite{Bernardeau2002}
\be
B^s(\bfk_1,\bfk_2,\bfk_3) & = & 2 D_+^4 P_{L0}(k_1) P_{L0}(k_2) Z_1(\bfk_1) Z_1(\bfk_2)
\nonumber \\
&& \times \, Z_2(\bfk_1,\bfk_2) + 2 \; \rm{perm.} \; ,
\label{Bs-1loop}
\ee
where ``2 perm.'' stands for two other terms that are obtained from permutations
over the indices $\{1,2,3\}$, and the kernels $Z_1$ and $Z_2$ are given by
\be
Z_1(\bfk) = 1 + f \mu^2 ,
\label{Z1-def}
\ee
and
\be
Z_2(\bfk_1,\bfk_2) & = & \frac{5+3f\mu^2}{7} + \frac{1+f\mu^2}{2} 
\left( \frac{k_1}{k_2} + \frac{k_2}{k_1} \right) \frac{\bfk_1\cdot\bfk_2}{k_1 k_2} \nonumber \\
&& + \frac{2+4f\mu^2}{7} \left( \frac{\bfk_1\cdot\bfk_2}{k_1k_2} \right)^2 
+ \frac{f k \mu}{2} \nonumber \\
&& \times \; \left[ \frac{\mu_1}{k_1} (1+f\mu_2^2) + \frac{\mu_2}{k_2} (1+f\mu_1^2) \right] ,
\ee
where $\bfk=\bfk_1+\bfk_2$.
In the small-$k'$ limit we obtain
\be
B^s_{k'\rightarrow 0} & = & 2 D_+^4 P_{L0}(k') P_{L0}(k_1) Z_1(\bfk') Z_1(\bfk_1)
\nonumber \\
&& \times \, Z_2(\bfk',\bfk_1) + ( \bfk_1 \leftrightarrow \bfk_2 ) ,
\label{Bs-1loop-2}
\ee
with $\bfk_1= \bfk-\bfk'/2$ and $\bfk_2= -\bfk-\bfk'/2$.
Here we used the fact that $Z_2(\bfk_1,\bfk_2)$ vanishes
as $|\bfk_1+\bfk_2|^2$ for $|\bfk_1+\bfk_2|\rightarrow 0$, whereas 
$P_{L0}(k) \sim k^{n_s}$ with $n_s \lesssim 1$. 
[If this is not the case, that is, there is very little initial power 
on large scales, we must go back to the consistency relation in the form 
of Eq.(\ref{tCn0-1}) rather than Eq.(\ref{tCn0-2}). However, this is not necessary
in realistic models.]
Expanding the various terms over $k'$, as
\be
P_{L0}(k_1) = P_{L0}(k) - \frac{\bfk\cdot\bfk'}{2\bfk} \frac{\dd P_{L0}}{\dd k}(k) + .. ,
\ee
\be
Z_1(\bfk_1) = 1 + f \mu^2 - f \frac{k'}{k} \mu\mu' + f\mu^2 \frac{\bfk\cdot\bfk'}{k^2} + .. ,
\ee
\be
Z_2(\bfk',\bfk_1) & \! = \! & \frac{13 \!+\! 19f\mu^2}{28} + \frac{4 \!+\! f\mu^2}{14} 
\left(\! \frac{\bfk\cdot\bfk'}{k k'} \!\right)^2 + (1 \!+\! f\mu^2) \nonumber \\
&& \hspace{-1.5cm} \times \; \left[ \frac{f\mu'^2}{4} + \frac{f\mu\mu'}{2} \frac{\bfk\cdot\bfk'}{k k'} 
+ \frac{\bfk\cdot\bfk'}{2k'^2} + \frac{f\mu\mu'}{2} \frac{k}{k'} \right] + .. , \;\;
\ee
substituting into Eq.(\ref{Bs-1loop-2}), and integrating over the angles of $\bfk'$, we obtain
\be
\int \frac{\dd \Omega_{\bfk'}}{4\pi} \frac{B^s_{k'\rightarrow 0}}{1+f\mu'^2} & = &  
P_L(k') P_L(k) (1+f\mu^2) \left[ \frac{47}{21} + \frac{f}{3} \right. \nonumber \\
&& \left. + \frac{73 f\mu^2}{21} - \frac{f^2\mu^2}{3} + \frac{4 f^2\mu^4}{3} \right] \nonumber \\
&& - P_L(k') \frac{\dd P_L(k)}{\dd\ln k} \frac{(1+f\mu^2)^3}{3} .
\label{B-spec-tree-1}
\ee

On the other hand, the right-hand side of Eq.(\ref{eq:bispec}) reads at the same order
over $P_L$ as
\be
\hspace{-0.cm} \int \frac{\dd \Omega_{\bfk'}}{4\pi} \frac{B^s_{k'\rightarrow 0}}
{1+f\mu'^2} & = & P_L(k') \left[1+ \frac{f}{3} + \frac{13}{21}\frac{\partial}{\partial\ln D_+} 
\right. \nonumber \\
&& \hspace{-2.5cm} + \left( \frac{13}{7} + f \right) \frac{f}{3} \frac{\partial}{\partial f} 
- \frac{1+f\mu^2}{3}\frac{\partial}{\partial\ln k} \nonumber \\
&& \hspace{-2.5cm} \left. - \frac{2 f \mu^2}{3} (1-\mu^2) \frac{\partial}{\partial \mu^2} \right] 
D_+^2 P_{L0}(k) (1+f\mu^2)^2 . \;\;
\label{bispec-tree-2}
\ee
Collecting the various terms we can check that we recover Eq.(\ref{B-spec-tree-1}).

Therefore, we have checked the angular-averaged redshift-space consistency relation
(\ref{tCn0-3}) for the bispectrum, at leading order of perturbation theory,
within the approximate symmetry $\Omega_{\rm m}/f^2 \simeq 1$ discussed in
Sec.~\ref{sec:approx-sym}.
In this explicit check, the use of this approximate symmetry appears at the level of
the expression (\ref{Bs-1loop}) of the bispectrum, which only involves the linear 
growing mode $D_+$ and the factor $f$ as functions of time and cosmology.
An exact calculation would give factors that show new but weak dependencies 
on time and cosmology (and that are unity for the Einstein-de Sitter case) \cite{Bernardeau2002}.
These deviations from Eq.(\ref{Bs-1loop}) are usually neglected 
[for instance, when the cosmological constant is zero, they were shown to be well 
approximated by factors like $(\Omega_{\rm m}^{-2/63}-1)$ that are very small over the range
of interest \cite{Bouchet1992}].

For future use in Sec.~\ref{sec:full-form} below, in terms of the angular monopole and
quadrupole, Eq.(\ref{B-spec-tree-1}) gives at lowest order of perturbation theory:
\be
\hspace{-0.cm} \int_{-1}^{1} \!\! \frac{\dd\mu}{2} \int \! \frac{\dd \Omega_{\bfk'}}{4\pi} 
\frac{B^s_{k'\rightarrow 0}}{1+f\mu'^2} & \! = \! & P_L(k') \biggl \lbrace \left( 235+235 f+101 f^2
\right. \nonumber \\
&& \hspace{-4cm} \left. +13 f^3 \right) \frac{P_L(k)}{105} 
- \frac{35+35 f+21 f^2+5 f^3}{105} \frac{\dd P_L(k)}{\dd\ln k} \biggl \rbrace
\label{bispec-tree-L0}
\ee
and
\be
\hspace{-0.cm} \frac{5}{2} \int_{-1}^{1} \!\! \dd\mu L_2(\mu) \int \! 
\frac{\dd \Omega_{\bfk'}}{4\pi} \frac{B^s_{k'\rightarrow 0}}{1 \!+\! f\mu'^2} & \! = \! & 
P_L(k') \biggl \lbrace \frac{4 f}{441} \left( 420 \!+\! 303 f \right. \nonumber \\
&& \hspace{-4.5cm} \left. +49 f^2 \right) P_L(k)
- \frac{2f}{63} \left( 21+18 f+5 f^2 \right) \frac{\dd P_L(k)}{\dd\ln k} \biggl \rbrace .
\label{bispec-tree-L2}
\ee

\subsection{1D nonlinear check}
\label{sec:check-1D}

The explicit check presented in Sec.~\ref{sec:perturbative} only applies up to the
lowest order of perturbation theory. Because the goal of the consistency relations 
is precisely to go beyond low-order perturbation theory, it is useful to
obtain a fully nonlinear check. This is possible in one dimension, where the
Zel'dovich solution \cite{ZelDovich1970} becomes exact (before shell crossing)
and all quantities can be explicitly computed.
Because of the change of dimensionality, we also need to rederive the 1D form of
the consistency relations. We present the details of our computations in
App.~\ref{sec:1D-example}, and only give the main steps in this section.

In the 1D case, the redshift-space coordinate (\ref{s-def-1}) now reads as
\be
s = x + \frac{v}{\dot{a}} = x + f u ,
\label{s-def-1D}
\ee
where $u$ is the rescaled peculiar velocity defined in Eq.(\ref{eta-1D}), in a fashion
similar to Eq.(\ref{eta-3D}), and the redshift-space density contrast (\ref{tdelta-s-def}) 
now writes as
\be
k \neq 0 : \;\;\; \tdelta^s(k,t) = \int \frac{\dd q}{2\pi} \; e^{-\ii k s(q,t)} ,
\label{tdelta-s-def-1D}
\ee
where we again discarded a Dirac term that does not contribute for $k\neq 0$
and $q$ is the Lagrangian coordinate of the particles.

As in the 3D case, to derive the 1D consistency relations we consider two universes
with close cosmological parameters and expansion rates, $a'(t) = [1-\epsilon(t)] a(t)$.
Again, from the ``1D Friedmann equations'' we find that $\epsilon(t) = \epsilon_0 D_+(t)$.
Next, a uniform overdensity $\Delta_{L0}$ can be absorbed by a change of frame,
with $\epsilon_0=\Delta \delta_{L0}$.
Then, to obtain the consistency relations, we need the impact of the large-scale overdensity
$\Delta_{L0}$ on small-scale structures, which at lowest order is given by the dependence
of the small-scale density contrast $\tdelta^s(k)$ on $\epsilon_0$.
As shown in App.~\ref{sec:1D-background-perturbation}, this reads as
\be
\frac{\partial \tdelta^s(k,t)}{\partial\epsilon_0} & = & 
D_+^2 \frac{\partial \tdelta^s}{\partial D_+} 
+ f D_+ (1+f) \frac{\partial \tdelta^s}{\partial f} \nonumber \\
&& - (1+f) D_+ k \frac{\partial\tdelta^s}{\partial k} .
\label{d-delta-d-eps0-1D-2}
\ee
As expected, this takes the same form as the 3D result (\ref{d-deltak-d-eps0-2}),
up to some changes of numerical coefficients.
This leads to the equal-time redshift-space consistency relations 
(see App.~\ref{sec:1D-consistency-relations})
\be
\biggl \langle \frac{\tdelta^s(k')}{1+f} \tdelta^s(k_1) .. 
\tdelta^s(k_n) \biggl \rangle'_{k'\rightarrow 0} =  P_L(k') \biggl [ 1+f + \frac{\partial}{\partial\ln D_+} 
\nonumber \\
&& \hspace{-8cm} + (1+f) f \frac{\partial}{\partial f}  - (1+f) \sum_{i=1}^n k_i \frac{\partial}{\partial k_i} \biggl ]
\langle \tdelta^s(k_1) .. \tdelta^s(k_n) \rangle' . \nonumber \\
&&
\label{tCn-2-1D}
\ee
Here we no longer need to average over the directions of the large-scale wave number $k'$,
because at equal times the leading-order contribution associated with the uniform displacement
of small-scale structures by large-scale modes vanishes
\cite{Kehagias2013,Peloso2013,Creminelli2013,Kehagias2014,Peloso2014,Creminelli2014,Valageas:2014aa}.
Indeed, because of statistical homogeneity and isotropy, equal-time polyspectra are invariant
through uniform translations and cannot probe uniform displacements.
Therefore, 1D equal-time statistics directly probe the next-to-leading order contribution
(\ref{tCn-2-1D}), which truly measures the impact of large-scale modes on the growth of small-scale
structures.

In the 1D case, the Zel'dovich approximation is exact until shell crossing 
\cite{ZelDovich1970,Valageas:2014ab} and it yields for the
redshift-space nonlinear density contrast (\ref{tdelta-s-def-1D}) the expression
(see App.~\ref{sec:Zel'dovich})
\be
\tdelta^s(k,t) = \int \frac{\dd q}{2\pi} \; e^{-\ii k q + k (1+f) D_+ \int \frac{\dd k'}{k'} 
e^{\ii k' q} \tdelta_{L0}(k') } . \;\;
\label{tdelta-s-Zel}
\ee
The expression (\ref{tdelta-s-Zel}) is exact at all orders of perturbation theory, but it no longer
holds after shell crossing (which is a nonperturbative effect).
On the other hand, we can define a 1D toy model by setting particle trajectories as equal to
Eq.(\ref{Zel-def}). This system is no longer identified with a 1D gravitational system, and it
only coincides with the latter in the perturbative regime, but it remains well defined and given
by Eqs.(\ref{Zel-def}) and (\ref{tdelta-s-Zel}) in the nonperturbative shell-crossing regime.

Then, using the expression (\ref{tdelta-s-Zel}) we can explicitly check the 1D consistency
relations (\ref{tCn-2-1D}).
We present in App.~\ref{sec:1D-check} two different checks.
First, in App.~\ref{sec:Impact-1D}, we check Eq.(\ref{d-delta-d-eps0-1D-2}) by 
explicitly computing the impact on the nonlinear density contrast (\ref{tdelta-s-Zel})
of a small change $\Delta\delta_{L0}$ to the initial conditions.
Second, in App.~\ref{sec:Explicit-1D}, we directly check the consistency
relations (\ref{tCn-2-1D}) by explicitly computing the correlations
$\langle \tdelta_L(k') \tdelta^s(k_1) .. \tdelta^s(k_n) \rangle'_{k'\rightarrow 0}$ and
$\langle \tdelta^s(k_1) .. \tdelta^s(k_n) \rangle'$ and verifying that they satisfy
Eq.(\ref{tCn-2-1D}).

These two different checks allow us to check both the reasoning that leads to the
consistency relations, through the intermediate result (\ref{d-delta-d-eps0-1D-2}),
and the final expression of these relations.
They also explicitly show that they are not restricted to the perturbative regime.
In particular, they extend beyond shell crossing, as seen from the toy model
defined by the explicit expression (\ref{tdelta-s-Zel}) (i.e., where one defines the
system by the Zel'dovich dynamics, even beyond shell crossing, without further
reference to gravity).

As for the real-space consistency relations \cite{Valageas:2014ab}, it happens that
in this 1D model (\ref{tdelta-s-Zel}) the 1D consistency relations (\ref{tCn-2-1D}) are actually
exact, that is, they do not rely on the approximation $\kappa \simeq \kappa_0$, where
$\kappa$ defined in Eq.(\ref{kappa-def}) plays the role of the 3D factor
$\Omega_{\rm m}/f^2$ encountered in Eqs.(\ref{cont-2})-(\ref{Poisson-2}).
This is because the redshift-space density contrast (\ref{tdelta-s-Zel})
truly only depends on cosmology and time through the two factors $D_+$ and $f$,
even at nonlinear order. In contrast, in the 3D gravitational case, beyond linear order
new functions of cosmology and time appear (for cosmologies that depart from
the Einstein-de Sitter case) and they can only be reduced to powers of $D_+$ and $f$
within the approximation $\Omega_{\rm m}/f^2 \simeq 1$.
On the other hand, if we consider the actual 1D gravitational dynamics even beyond
shell crossing, where it deviates from the expression (\ref{tdelta-s-Zel}), then
the 1D consistency relations (\ref{tCn-2-1D}) are only approximate in the nonperturbative
regime, as they rely on the approximation $\kappa \simeq \kappa_0$, while remaining
exact at all perturbative orders.

Unfortunately, it is not easy to build 3D analytical models that can be explicitly solved
and suit our purposes.
The 3D Zel'dovich approximation again provides a simple model for the formation
of large-scale structures and the cosmic web. However, it cannot suit our purposes
because it does not apply to the dynamics of the 3D background universe itself.
Indeed, as can be seen from their derivation in Sec.~\ref{sec:angular}, 
the consistency relations precisely derive from the fact that a large-scale almost uniform
density perturbation can be seen as a local change of the cosmological parameters
(i.e., the background density).
This is also apparent through the fact that the deviation $\epsilon(t)$ between the two
nearby universes (\ref{eps-def}) obeys the same evolution equation (\ref{Dlin-1})
as the linear growing mode of local density perturbations.
This is no longer possible for the 3D Zel'dovich approximation, which is not an exact
solution and cannot be extended to the Hubble flow itself.
In contrast, in the 1D universe the Zel'dovich approximation is actually exact (before
shell crossing) and it applies both at the level of the background and of the density
perturbations.
An alternative dynamics, which is exact at the background level and provides
analytical results on small nonlinear scales, is the spherical collapse model.
However, this yields a very different density field than the actual one, as there is
a single central density fluctuation that breaks statistical homogeneity and
density correlations are no longer invariant through translations.
Therefore, although it should be possible to obtain some consistency relations
for this model, they would have a rather different form and this 1D spherical model
would be even farther from the actual universe than the 1D statistically homogeneous
model studied in this section.

\section{Simulations}
\label{sec:simulations}

The angular-averaged consistency relations (\ref{tCn0-1})-(\ref{tCn0-3}) 
are valid at all orders of perturbation theory and also beyond the
perturbative regime, including shell-crossing effects, within the accuracy
of the approximation $\Omega_{\rm m}/f^2 \simeq 1$ (and as long as gravity is the
dominant process). We have explicitly confirmed them either perturbatively
or nonperturbatively, but the latter is limited to the one-dimensional case. 

It would thus be of great importance to further check these relations in three dimensions nonperturbatively. 
We here exploit a series of $N$-body simulations for this purpose.
As can be seen in the following, they are also useful to understand the possible breakdown of the relations and 
test the validity of the ansatz employed in the measurement in practical situations. 
We first summarize how we can evaluate the derivative terms
in the consistency relations. We then present the numerical results for the bispectrum together with a
brief description of the simulations themselves.

\subsection{Derivatives from numerical simulations}
\label{sec:deriv}

The consistency relation (\ref{eq:bispec}) involves derivatives with respect to $D_+$ and
$f$. They can be obtained at once within the framework of an explicit analytic model
for the matter density polyspectra. However, in this paper we do not use these relations
to check a specific analytical model. Instead, we wish to use numerical simulations
to test these relations (which are only approximate because of the approximation 
$\Omega_{\rm m}/f^2 \simeq 1$).
Nevertheless, we can also measure separately the derivatives with
respect to $D_+$ and $f$ from the simulations.

The redshift-space coordinate $\bfs$ can be written in terms of the comoving coordinate
$\bfx$ and peculiar velocity $\bfv$ as in Eq.(\ref{s-def}).
As explained in Sec.~\ref{sec:approx-sym}, within the approximation 
$\Omega_{\rm m}/f^2 \simeq 1$
that is used to derive the consistency relations, all time dependence can be absorbed in
the linear growing mode $D_+(t)$ with the change of variables (\ref{eta-3D}).
This means that the fields $\{\delta,\bfu,\varphi\}$ are only functions of time through $D_+$,
as well as the displacement field $\Psi(\bfq,t) = \bfx -\bfq$, where $\bfq$ is the Lagrangian coordinate
of the particles.
Thus, for a given realization defined by the linear density field $\delta_{L0}(\bfq)$
(normalized today or at the initial time of the simulation),
the redshift-space coordinate $\bfs$ depends on the functions $D_+(t)$ and $f(t)$
as
\be
\bfs(\bfq,t) = \bfx(\bfq,D_+) + f u_r(\bfq,D_+) \, \bfe_r .
\ee
Therefore, a small change $\Delta f$ of the factor $f$ corresponds to a change of the
redshift-space coordinate $\bfs(\bfq)$ of the particles given by:
\be
f \rightarrow f+\Delta f : \;\; \bfs \rightarrow \bfs + \Delta f \; u_r \, \bfe_r 
= \bfs + \Delta f \frac{v_r}{\dot{a}f} \, \bfe_r .
\label{s-Delta-f}
\ee

On the other hand, from the equations of motion (\ref{cont-2})-(\ref{Poisson-2}), a change
$\Delta \ln D_+$ of the linear growing mode leads to a change of the particle velocities and
coordinates
\be
\hspace{-0.5cm} \ln D_+ \rightarrow \ln D_+ + \Delta\ln D_+ : &&  
\bfx \rightarrow \bfx + \Delta\ln D_+ \; \bfu , \nonumber \\
&& \hspace{-3.5cm} \bfu \rightarrow \bfu - \Delta\ln D_+ \; \left[ 
\left( \frac{3\Omega_{\rm m}}{2 f^2} -1 \right) \bfu + \nabla\varphi \right] ,
\ee
whence,
\be
\bfs & \rightarrow & \bfs + \Delta\ln D_+ \left[ \bfu - f \left( 
\left( \frac{3\Omega_{\rm m}}{2 f^2} -1 \right) u_r 
+ \frac{\partial\varphi}{\partial r} \right) \bfe_r \right] \nonumber \\
&& \hspace{-0.4cm} = \bfs + \Delta\ln D_+ \left[ \frac{\bfv}{\dot{a}f} - \left(\! 
\left(\! \frac{3\Omega_{\rm m}}{2 f^2} -1 \! \right) \frac{v_r}{\dot{a}} 
+ \frac{1}{\dot{a}^2f} \frac{\partial\phi}{\partial r} \! \right) \bfe_r \right]  \! . \nonumber \\
&&
\label{s-Delta-lnD}
\ee

Thus, to obtain the partial derivative of the power spectrum with respect to $f$ or $\ln D_+$,
we modify the particle redshift-space coordinates by Eqs.(\ref{s-Delta-f}) or (\ref{s-Delta-lnD}),
for a small value of $\Delta f$ or $\Delta\ln D_+$, and we compute the associated power spectrum.
Taking the difference from the initial power spectrum and dividing by $\Delta f$ or $\Delta\ln D_+$
gives a numerical estimate of $\partial P^s/\partial f$ or $\partial P^s/\partial \ln D_+$.

\subsection{Numerical results}
\label{sec:numerical}

We are now in a position to present the consistency relations measured from simulations.
Before that, let us briefly describe the simulations used here.
They are the ones performed in \cite{Taruya2012}.
Employing $1024^3$ dark matter particles in a periodic cube of $(2048\,h^{-1}\mathrm{Mpc})^3$, the gravitational dynamics
was solved by a public simulation code \texttt{Gadget2} \cite{Springel2005} starting from an initial condition set at $z=15$
by solving second-order Lagrangian perturbation theory \cite{Scoccimarro1998,Crocce2006,Nishimichi2009}.
The cosmological model used was a flat-$\Lambda$CDM model consistent with the five-year observation of the WMAP
satellite \cite{Komatsu2003}: $\Omega_\mathrm{m}=0.279$, $\Omega_\mathrm{b}=0.165\Omega_\mathrm{m}$, 
$h=0.701$, $A_\mathrm{s}=2.49\times10^{-9}$ and $n_\mathrm{s}=0.96$ at $k_0 = 0.002\mathrm{Mpc}^{-1}$.
This whole process was repeated $60$ times with the initial random phases varied to have a large ensemble of 
random realizations.

The consistency relations have already been examined and presented in real space in \cite{Nishimichi2014}.
There it was found that the relation was recovered within the numerical accuracy at $z=1$, while discrepancy of
several percent level was found at $z=0.35$. It was further discussed that this is presumably due to 
the breakdown of the approximation $\Omega_\mathrm{m}/f^2 \simeq 1$; we could indeed confirm that the relations 
better hold in supplementary simulations done in the EdS background, but with exactly the same initial perturbations. 
We focus here on the lower redshift, $z=0.35$, at which the consistency relations are 
the most nontrivial.

\subsubsection{Full consistency relations}
\label{sec:full-form}

We first consider the redshift-space consistency relations in their full form
(\ref{eq:bispec-ell-0})-(\ref{eq:bispec-ell-2}), with both derivative operators
$\partial/\partial D_+$ and $\partial/\partial f$.

Having already presented the methods we employ to measure the derivative terms in the previous subsection, 
the post-processing for the simulation outputs is exactly the same as in \cite{Nishimichi2014} except that 
we now consider the particle positions in redshift space. The matter density field is constructed with the Cloud-in-Cells
(CIC) interpolation on $1024^3$ mesh cells and subsequent computations are based on the fast Fourier transform.
The change in the particle coordinates corresponding to a slight change in $\ln D_+$ is also computed based 
on the calculation on the same mesh cells for $\partial \phi / \partial r$ and then interpolated to the positions of particles
using the CIC kernel (see Eq.~\ref{s-Delta-lnD}).

\begin{figure}[!ht]
   \centering
   \includegraphics[width=8.5cm]{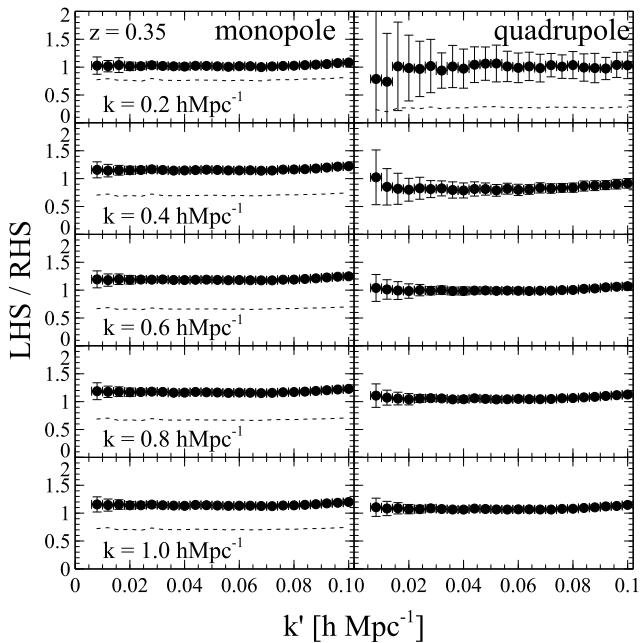} 
   \caption{Consistency-relation ratio for the redshift-space bispectrum from $N$-body simulations. The dashed lines show the ratio of the measured bispectrum to its tree-order
 predictions (\ref{bispec-tree-L0}) and (\ref{bispec-tree-L2}).}
   \label{fig:ratio_raw}
\end{figure}

The monopole and the quadrupole moments of the relation for the bispectra, Eqs.(\ref{eq:bispec-ell-0}) and 
(\ref{eq:bispec-ell-2}), are respectively shown in the left and the right panel of Fig.~\ref{fig:ratio_raw}.
In each panel, we fix the value of the hard wave mode $k$, and plot the ratio of the two sides as a function of the soft mode $k'$.
The error bars are estimated based on the scatter among the $60$ independent realizations.
They thus correspond to the error level expected for an ideal survey with a volume of $\sim 8\,h^{-3}\mathrm{Gpc}^3$ when
we can ignore the shot noise contamination. Overall, the ratio is close to unity for both monopole and quadrupole.
From this figure, we basically confirm the relations at the nonperturbative level in the three dimensional dynamics.

The dashed lines in Fig.~\ref{fig:ratio_raw} show the ratio of the measured bispectrum 
to its tree-order predictions (\ref{bispec-tree-L0}) and (\ref{bispec-tree-L2}).
For the monopole, this lowest-order perturbative prediction fares reasonably well
as it only underestimates the nonlinear results by $30\%$, on these scales.
However, it is already less accurate than our result (\ref{eq:bispec-ell-0}), which
takes into account higher-order and nonperturbative nonlinear corrections
(at the price of the approximation $\Omega_{\rm m}/f^2 \simeq 1$).
For the quadrupole, the lowest-order perturbative prediction does not appear in the 
panels at $k \geq 0.4 h$Mpc$^{-1}$ because in these cases it is out of range and 
actually gives the wrong sign. This change of sign is likely due to the fingers-of-god effect,
which is not captured by perturbation theory.
Indeed, it is well known that higher-order multipoles are increasingly sensitive to
small-scale nonlinear contributions, as finger-of-god effects impart a strong angular 
dependence to the bispectrum \cite{Scoccimarro1999}.
In contrast, our result (\ref{eq:bispec-ell-2}) remains consistent with the simulation data
within $20\%$. This shows that we test the consistency relations in a nontrivial regime,
beyond the reach of standard perturbation theory.
Thus, the trade-off between the error introduced by the approximate symmetry of 
Sec.~\ref{sec:approx-sym} and the advantage of taking into account all nonlinear 
contributions, at both perturbative and nonperturbative levels, is beneficial.
This is particularly true for complex statistics such as the redshift-space quadrupole
that are very sensitive to small-scale highly-nonlinear effects, which are difficult to include
in analytical modelings.

However, when we look into each panel more closely, we can find that the data points are slightly off from unity.
For the monopole moment, the ratio tends to be larger than unity at $k\simgt 0.4\,h\mathrm{Mpc}^{-1}$.
On the other hand, unity is within the statistical error level for the quadrupole moment, though the central values
are larger (smaller) than unity on $k\simlt 0.4\,h\mathrm{Mpc}^{-1}$ ($k\simgt 0.8\,h\mathrm{Mpc}^{-1}$).
In most of the cases, the deviation from unity is at most $20\%$, and this is meaningful only when we measure the
ratio very precisely; an ideal survey with a volume of $\sim 8\,h^{-3}\mathrm{Gpc}^3$ can detect the deviation from
unity only for the monopole moment on small scales.

These deviations are somewhat greater than those found in \cite{Nishimichi2014}
in real-space, which only reached $7\%$ at $k=1h$Mpc$^{-1}$.
This is not surprising, because it is well known that redshift-space statistics are more
sensitive to small nonlinear scales, for instance through the fingers of god effect,
and low-order perturbation theory has a smaller range of validity.
Then, we can expect a greater violation of the redshift-space consistency relations
because the breakdown of the approximation $\Omega_{\rm m}/f^2 \simeq 1$
has a stronger impact on higher perturbative orders.
Indeed, absorbing the time and cosmological dependence by $D_+(t)$ and $f(t)$ is 
exact at linear order whereas higher orders involve new functions $D_+^{(n)}(t)$ 
that are not exactly equal to $D_+(t)^n$ \cite{Bernardeau2002} and the discrepancies 
may cumulate in the nonlinear regime.

\begin{figure}[!ht]
   \centering
   \includegraphics[width=8.5cm]{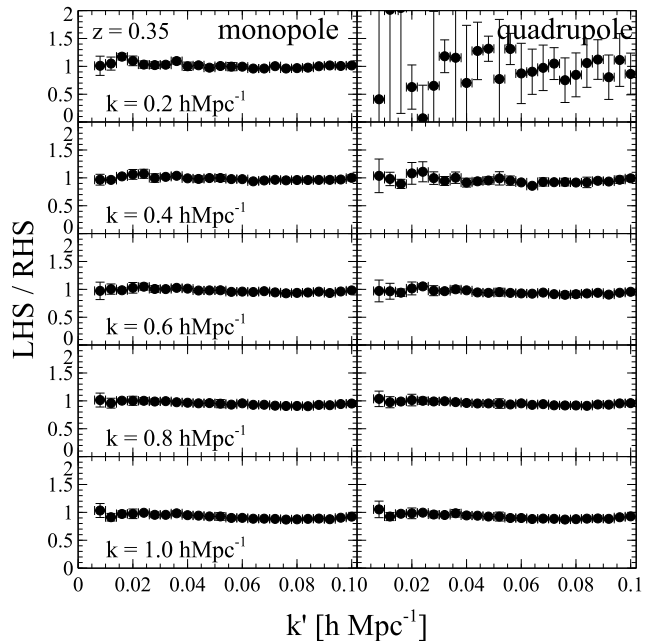} 
   \caption{Same as Fig.~\ref{fig:ratio_raw}, but for the simulations performed in EdS background expansion.}
   \label{fig:ratio_EdS}
\end{figure}

We then work on the supplemental simulations done in the EdS background, to understand 
the cause of this small discrepancy, just as in our previous real-space paper 
\cite{Nishimichi2014}. Note that our consistency relations in an Einstein-de Sitter 
cosmology also involve the approximate symmetry described in Sec.~\ref{sec:approx-sym},
even though $\Omega_{\rm m}/f^2=1$ in the EdS background. Indeed, what matters
is not that $\Omega_{\rm m}/f^2$ be unity in the reference cosmology, but that
$\Omega_{\rm m}/f^2$ remain (approximately) constant as we vary the background
curvature around the reference cosmology.
Nevertheless, the comparison between EdS and $\Lambda$CDM results provides
a simple estimate of the impact of our approximation, because the difference between
these two cosmologies arises from the change of reference point along the
$\Omega_{\rm m}/f^2$ curve.

The results from four realizations of such simulations are shown in Fig.~\ref{fig:ratio_EdS}. 
Although the scatter of the data points are larger than in Fig.~\ref{fig:ratio_raw}, the 
systematic departure from unity in the previous figure is clearly reduced. We thus conclude 
that the small violation of the consistency relations for the bispectrum can be explained 
by the breakdown of the approximation $\Omega_\mathrm{m}/f^2 \simeq 1$
(more precisely, of constant $\Omega_\mathrm{m}/f^2$ for nearby background
curvatures), in agreement with the discussions above.

\subsubsection{$f\mu^2$ ansatz and reduction to $\partial/\partial D_+$ operator}
\label{sec:ansatz1}

\begin{figure}[!ht]
   \centering
   \includegraphics[width=8.5cm]{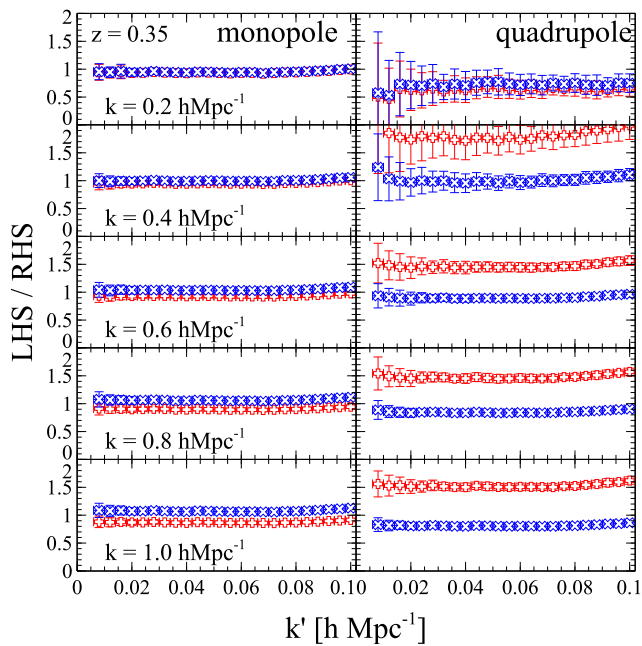} 
   \caption{Same as Fig.~\ref{fig:ratio_raw}, but for Eqs.~(\ref{eq:bispec-ell-0-f1}) and (\ref{eq:bispec-ell-2-f1}) after applying the ansatz to rewrite the $f$-derivative terms.
   Blue crosses and red plus symbols respectively show the ratio truncated at $P_2^s$ and $P_4^s$.}
   \label{fig:ratio_app1}
\end{figure}

Now, we come back to the original $\Lambda$CDM simulations and apply the ansatz that all 
the $f$ and $\mu$ dependences appear through the combination $f\mu^2$. 
This allows us to replace the $f$-derivatives, $\partial/\partial f$, by $\mu^2$-derivatives,
$\partial/\partial\mu^2$, as in Eq.(\ref{df-dmu2-def}).
This gives the approximated consistency relations for the bispectrum, 
Eqs.~(\ref{eq:bispec-ell-0-f1}) and (\ref{eq:bispec-ell-2-f1}), 
respectively for the monopole and quadrupole moment, which we plot 
in Fig.~\ref{fig:ratio_app1}.
In contrast with the exact form of the consistency relations, given by 
Eqs.(\ref{eq:bispec-ell-0}) and (\ref{eq:bispec-ell-2}) and displayed in
Fig.~\ref{fig:ratio_raw}, the right-hand side now involves an infinite summation over
all Legendre multipoles of the redshift-space power spectrum.
Here, we truncate these series at order $P_2^s$ (crosses) or $P_4^s$ (pluses).

The difference between the two symbols is negligible at $k=0.2$ and 
$0.4\,h\mathrm{Mpc}^{-1}$ for the monopole and $k=0.2\,h\mathrm{Mpc}^{-1}$ for the 
quadrupole moment, where the ratio itself is roughly consistent with unity.
As we move to smaller scales, the two symbols become more distinct. In those cases,
adding the higher-order term (i.e., $P_4^s$) does not help to restore the relations, 
suggesting that the ansatz is not a good approximation at the corresponding scales.
The plot suggests that the quadrupole moment is more sensitive to the higher-order term 
and thus the ansatz works less accurately than for the monopole moment.
This is naturally expected since the quadrupole moment is impacted more strongly by 
higher-order corrections [see e.g. \cite{Taruya10}, where we can see how much 
higher-order perturbative corrections affect the first two moments.
These corrections have terms $\mu^{2m}f^n$, where $m$ and $n$ can be different.].

\subsubsection{$f\mu^2$ ansatz and further reduction to $\partial/\partial a$ operator}
\label{sec:ansatz2}

\begin{figure}[!ht]
   \centering
   \includegraphics[width=8.5cm]{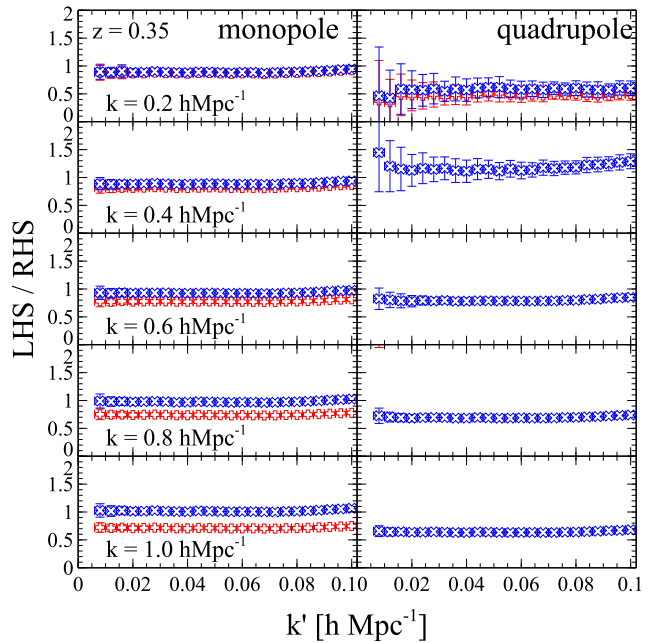} 
   \caption{Same as Fig.~\ref{fig:ratio_app1}, but for Eqs.~(\ref{eq:bispec-ell-0-f2}) and (\ref{eq:bispec-ell-2-f2}) after applying the ansatz once again to convert the 
   $D_+$-derivative terms into the $a$-derivatives.}
   \label{fig:ratio_app2}
\end{figure}

The situation is basically the same after we further apply the ansatz to replace the 
derivative with respect to $D_+$ by a derivative with respect to time or the scale
factor, as in Eq.(\ref{D+_a}). As compared with the form displayed in 
Fig.~\ref{fig:ratio_app1}, this involves an additional approximation, which relies
on the same $f\mu^2$-ansatz, because the full time derivative, or scale-factor
derivative, $\partial/\partial a$, combines both theoretical derivatives $\partial/\partial
D_+$ and $\partial/\partial f$. Therefore, to replace the operator 
$\partial/\partial D_+$ by $\partial/\partial a$ we must once again use the
$f\mu^2$-ansatz to remove the new $\partial/\partial f$ terms generated by the change
of variable from $D_+$ to $a$.

Figure~\ref{fig:ratio_app2} shows the results of Eqs.~(\ref{eq:bispec-ell-0-f2}) and 
(\ref{eq:bispec-ell-2-f2}).
The plus symbols are now out of the plotted range for the quadrupole moment on small 
scales.
The relation for the monopole moment is more robust against this approximation ansatz 
on large scales, especially at $k=0.2\,h\mathrm{Mpc}^{-1}$, and we can safely apply the 
consistency relation here in the simplified form (\ref{eq:bispec-ell-0-f2}).
Except for this case, the ratio is affected significantly by the ansatz and the order at which 
we truncate the infinite summation on the right-hand side. 
Nevertheless, we note that by truncating at order $P_2^s$ we obtain a good
agreement, better than $20\%$ up to $k=1h$Mpc$^{-1}$, for the monopole.
For the quadrupole, the deviation can reach up to $40\%$.

Therefore, even with the current ansatz, we can still examine how the ratio behaves
in the observations and compare it with the simulation results.
Since the data points obtained with the truncation at order $P_2^s$ (i.e., cross symbols) 
are less noisy and moreover stay around unity after applying the ansatz, the easiest check
of the true gravitational dynamics is to apply the same ansatz and truncate the
moments at this order.
We would need a more involved ansatz for the estimation of the derivative terms from 
observations to extend the applicable range of these consistency relations, and we leave 
this to a future study.

\section{Summary}
\label{sec:summary}

In this paper, we have generalized the equal-time angular-averaged consistency relations 
for the cosmic density field originally developed in real space by \cite{Valageas:2014ab} 
to redshift space, in which the actual observations are taking place. 
These relations express the squeezed limit of $(n+1)$-point correlation functions or 
polyspectra, with $n$ small-scale modes (that can be in the nonlinear regime) and 
one large-scale mode (in the linear regime at a much larger scale than all other $n$ wave 
numbers), in terms of the $n$-point correlation of the small-scale modes.
These relations can be generalized to $(n+\ell)$ correlations, with $\ell$ large-scale
modes, as in \cite{Valageas:2014ab}, but we focused here on the case of one large-scale
mode.
The explicit forms that we have obtained rely on an approximate symmetry of the
dynamics, $\Omega_{\rm m}/f^2 \simeq 1$. However, within this approximation
they are valid at a fully nonlinear level. Thus, they hold at all orders of perturbation 
theory and also in the nonperturbative regime, beyond shell crossing. 
In particular, they include both the large-scale Kaiser effect \cite{Kaiser:1987aa}, 
associated with the infall of matter within large-scale gravitational wells, and the 
fingers-of-god effect \cite{Jackson1972}, associated with the virial motions inside 
collapsed halos.

We have found that, because the mapping from real to redshift space involves the velocity
component along the radial direction, the form of these consistency relations is slightly
more complex than in real space, as it involves two types of time derivatives. 
The first is a derivative with respect to the linear growing mode $D_+(t)$, which also 
appeared in the real-space case.
The second is a derivative with respect to the linear growth rate, 
$f(t)=\dd\ln D_+/\dd\ln a$. This differential operator, $\partial/\partial f$, did not appear in the 
real-space case and it arises from the scaling of the peculiar velocity field
(i.e., through the change of variable from $\bfv$ to $\bfu$, where $\bfu$ is the
rescaled velocity field that makes use of the approximate symmetry of the dynamics).
This feature makes it more difficult to use these relations for observations,
because at best we can only measure one time derivative, $\partial/\partial t$, 
which combines both $\partial/\partial D_+$ and $\partial/\partial f$, and we cannot 
measure these two derivatives separately. However, these relations can still be used 
to check analytical models or numerical simulations, where we can explicitly compute
these two derivatives. 

Next, we have tested these consistency relations both analytically and numerically.
First, at leading order of perturbation theory, we have checked the lowest-order 
consistency relation, which expresses the squeezed limit of the bispectrum in terms of
the nonlinear power spectrum of the small-scale modes.
Second, in a fully nonlinear and nonperturbative analysis, we have checked
all these consistency relations at all orders, in the simpler one-dimensional case,
where we can use the exact Zel'dovich solution of the dynamics.

We have also tested the lowest-order consistency relations, relating the nonlinear
bispectrum and power spectrum, with numerical simulations.
We find a reasonably good agreement at $z=0.35$.
Projecting the angular dependence of the redshift-space polyspectra onto Legendre
polynomials, we find a good agreement for the monopole up to 
$k \lesssim 0.4 h$Mpc$^{-1}$ and we detect a small deviation of at most $20\%$
for $k \leq 1 h$Mpc$^{-1}$. For the quadrupole, we do not detect significant deviations
(but the statistical error bars are slightly larger).
In the case of an Einstein de Sitter cosmology, we find that these deviations are greatly
reduced and our numerical data agree with theoretical predictions.
Therefore, the small deviations found in the $\Lambda$CDM cosmology can be explained
by the finite accuracy of the approximation $\Omega_{\rm m}/f^2 \simeq 1$.

The typical magnitude of these deviations is larger and extends over a wider 
wave number range than for the real-space consistency relations 
\cite{Nishimichi2014}. This is consistent 
with the observation that the nonlinearity in the cosmic velocity field is more sensitive to the 
local nonlinear structure on small scales, as small-scale effects can easily propagate 
to larger scales through the nonlinear mapping from real to redshift space. Indeed, it is 
well known that the perturbation-theory prediction of the matter power spectrum is more 
difficult in redshift space \cite{Bernardeau2002}. Then, because nonlinear effects are 
likely to amplify the breakdown of the approximation $\Omega_{\rm m}/f^2 \simeq 1$,
violations of the consistency relations due to breakdown of this approximate symmetry
are indeed expected to be greater in redshift space.

On the other hand, we find that our results for the bispectrum provide a significant
improvement over lowest-order perturbation theory, especially for the quadrupole
where the perturbative prediction even gives the wrong sign for $k \geq 0.4 h$Mpc$^{-1}$.
This is a signature of the strong impact of small-scale nonlinearities onto redshift-space
statistics, which are usually difficult to model analytically.
This shows that we test the consistency relations in a nontrivial regime, beyond low-order
perturbation theory. 
It also shows the interest of these nonlinear relations, as the inaccuracy introduced
by the approximate symmetry $\Omega_{\rm m}/f^2 \simeq 1$ is more than compensated
by the account of higher-order and nonperturbative nonlinear contributions.
This can be even more beneficial for statistics such as the redshift-space quadrupole
that are sensitive to highly nonlinear effects that are difficult to model.

To make the connection with observations, or to simplify the form of these consistency
relations, we also tested a simple ansatz that allows to remove the new operator
$\partial/\partial f$. This relies on the approximation that $f$ and $\mu^2$ only enter
the redshift-space power spectrum through the combination $f\mu^2$
(this is exact at linear order, in the Kaiser effect).
A first step allows us to remove the operator $\partial/\partial f$, which only leaves 
the operators $\partial/\partial D_+$ and $\partial/\partial k$ in multipole space.
The drawback is that the right-hand side of each consistency relation now involves
an infinite series over multipoles of all orders.
We find that this approximation gives rise to an additional source of discrepancy
between the numerical data and the analytic predictions, especially for the
quadrupole. Moreover, the result depends on the order at which we truncate the
multipole series in the right-hand side. It turns out that better results are obtained
when we truncate at the lowest order $P_2^s$.
This suggests that the $f\mu^2$-ansatz does not faithfully describe higher perturbative
or nonperturbative orders.

In a second step, we use once more the $f\mu^2$-ansatz to replace the operator
$\partial/\partial D_+$ by the full time-derivative, or scale-factor derivative
$\partial/\partial a$. As could be expected, we find that this further increase
the deviation from the numerical simulation data and the dependence on the
truncation order, especially for the quadrupole.

Nevertheless, we find that using the $f\mu^2$-ansatz and truncating at order
$P_2^s$ we obtain an agreement that is better than $20\%$ up to $k=1 h$Mpc$^{-1}$
for the monopole. For the quadrupole the deviation can reach $40\%$.
Although these limitations make the accessible range of these consistency relations 
rather narrow, we can still make use of them to predict the higher-order polyspectra.
There is substantial recent progress on the redshift-space clustering, but the calculations
are mostly limited to the power spectrum (e.g., 
\cite{Scoccimarro2004,Matsubara2008,Taruya10,Seljak11,Jennings11,Reid11}).
Using the relations developed here, one can compute, for instance, the angular-averaged 
bispectrum in redshift space by substituting these formulae for the power spectrum. Since 
the relations approximately hold down to very small scales, albeit not perfectly, they can be 
useful in estimating the covariance matrices of the redshift-space observables [we need 
the trispectrum to compute the matrix for the power spectrum].
Indeed, the accuracy required for the covariance matrices might not be as demanding 
as that for the spectra themselves.
A study along this line is undergoing now, and we wish to present the results elsewhere 
in near future.

A more complex issue is the problem of biasing, when we wish to connect
measures from galaxy surveys with theoretical predictions.
In principle, the approximate symmetry $\Omega_{\rm m}/f^2 \simeq 1$ that we used
to obtain explicit expressions no longer applies once we take into account galaxy
formation physics. Indeed, baryonic processes (cooling, star formation, ....) involve new
characteristic scales that explicitly break the symmetry of the dynamics. 
Then, a priori it is no longer possible to absorb the time-dependence of the dynamics
by a simple rescaling that only involves the linear growing mode.
Therefore, the relations we have obtained are not guaranteed to apply to the galaxy
density field itself, by making the naive replacement $\delta \rightarrow \delta_{\rm g}$.
One should rather use these relations as constraints on the matter density field, and
given a supplementary model that relates the galaxy field to the dark matter density
field, derives the consequences onto the galaxy density field.
Of course, this would depend on the model that is used to describe galaxy formation
and introduce an additional approximation.
We leave such a study for future works.

\begin{acknowledgments}

This work is supported in part by the French Agence Nationale de la Recherche
under Grant ANR-12-BS05-0002. T.N. is supported by Japan Society for the Promotion of Science (JSPS) Postdoctoral Fellowships for Research Abroad.
The numerical calculations in this work were carried out on Cray XC30 at Center for Computational Astrophysics, CfCA,
of National Astronomical Observatory of Japan.

\end{acknowledgments}

\appendix

\section{1D example}
\label{sec:1D-example}

As for the real-space consistency relations \cite{Valageas:2014ab}, 
it is interesting to check the redshift-space consistency relations (\ref{tCn0-1})-(\ref{tCn0-3})
obtained in this paper by using a simple one-dimensional example that can be exactly solved.
This is again provided by the Zel'dovich dynamics \cite{ZelDovich1970}, which is exact in 1D 
(before shell crossing).

\subsection{1D equations of motion}
\label{sec:1D-motion}

The 1D version of Eqs.(\ref{cont-1})-(\ref{Poisson-1}) reads as \cite{Valageas:2014ab}
\be
\frac{\partial\delta}{\partial t} + \frac{1}{a} \frac{\partial}{\partial x} [ (1+\delta) v ] = 0 ,  
\label{cont1D-2}
\ee
\be
\frac{\partial v}{\partial t} + H v + \frac{1}{a} v \frac{\partial v}{\partial x} = - \frac{1}{a} 
\frac{\partial \phi}{\partial x} ,  
\label{Euler1D-2}
\ee
\be
\frac{\partial^2\phi}{\partial x^2} = 4 \pi {\cal G}(t) \bar\rho a^2 \delta .  
\label{Poisson1D-2}
\ee
Here we generalized the 1D gravitational dynamics to the case of a time-dependent 
Newton's constant ${\cal G}(t)$. This allows us to obtain ever-expanding cosmologies,
similar to the 3D Einstein-de Sitter cosmology, for power-law cases
${\cal G}(t) \propto t^{\alpha}$ with $-2 < \alpha < -1$ [and $a(t) \propto t^{\alpha+2}$, 
$\bar\rho(t) \propto 1/a(t) \propto t^{-(\alpha+2)}$].

Linearizing these equations, we obtain the evolution equation of the linear modes of
the density contrast. It takes the same form as the usual 3D equation (\ref{Dlin-1}),
$\ddot{D} + 2 H(t) \dot{D} - 4\pi {\cal G}(t) \bar\rho(t) D = 0$,
but with a time-dependent Newton's constant and the 1D scale factor $a(t)$.

In a fashion similar to the change of variables (\ref{eta-3D}), we make the change of
variables
\be
\eta = \ln D_+ , \; v = \dot{a} f u , 
\; \phi = (\dot{a}f)^2 \varphi , \; \mbox{with} \; f=\frac{a\dot{D}_+}{\dot{a}D_+} , \;\;
\label{eta-1D}
\ee
and we obtain the rescaled equations of motion 
\be
\frac{\partial\delta}{\partial \eta} + \frac{\partial}{\partial x} [ (1+\delta) u ] = 0 ,  \label{cont1D-3}
\ee
\be
\frac{\partial u}{\partial \eta} + [ \kappa(t) - 1 ]  u + u \frac{\partial u}{\partial x} = 
- \frac{\partial \varphi}{\partial x} ,  
\label{Euler1D-3}
\ee
\be
\frac{\partial^2\varphi}{\partial x^2} = \kappa(t) \delta ,
\label{Poisson1D-3}
\ee
where we introduced the factor $\kappa(t)$ defined by
\be
\kappa(t) = 4\pi {\cal G}(t) \bar{\rho}(t) \frac{D_+(t)^2}{\dot{D}_+(t)^2} .
\label{kappa-def}
\ee
Thus, $\kappa(t)$ plays the role of the ratio $3\Omega_{\rm m}/(2f^2)$ encountered in
the 3D case in Eqs.(\ref{cont-2})-(\ref{Poisson-2}).
Then, the 3D approximation $\Omega_{\rm m}/f^2 \simeq 1$ used in the main text corresponds
in our 1D toy model to the approximation $\kappa \simeq \kappa_0$. That is, we
neglect the dependence of $\kappa$ on the cosmological parameters and time, and the
dependence on the background is fully contained in the change of variables (\ref{eta-1D}). 
[The generalization to the case of a time-dependent Newton's constant is not
important at a formal level, because it does not modify the form of the equations of
motion. However, it is necessary for this approximate symmetry to make practical sense,
so that we can find a regime where $\kappa$ is approximately constant.
This corresponds to cosmologies close to the Einstein-de Sitter-like expansion
$a(t) \propto t^{\alpha+2}$, in the case ${\cal G}(t) \propto t^{\alpha}$ with $-2<\alpha<-1$.] 

The fluid equations (\ref{cont1D-3})-(\ref{Poisson1D-3}) only apply to the 
single-stream regime, but we can again go beyond shell crossings by using the
equation of motion of trajectories, which reads as
\be
\frac{\partial^2 x}{\partial\eta^2} +  \left[ \kappa(t)  - 1 \right] 
\frac{\partial x}{\partial\eta} = -  \frac{\partial \varphi}{\partial x} ,
\label{traj1D-2}
\ee
where $\varphi$ is the rescaled gravitational potential (\ref{Poisson1D-3}).
This is the 1D version of Eq.(\ref{traj-2}) and it explicitly shows that particle
trajectories obey the same approximate symmetry, before and after shell crossings.

\subsection{1D background density perturbation}
\label{sec:1D-background-perturbation}

To derive the 1D consistency relations, we follow the method described in the main text 
for the 3D case, see also the Appendix in \cite{Valageas:2014ab}.
As in Eq.(\ref{eps-def}), we consider two universes with close cosmological
parameters, $a'(t) = a(t) [1-\epsilon(t) ]$ and $\bar\rho'(t) = \bar\rho(t) [ 1+\epsilon(t) ]$.
Substituting into the ``1D Friedmann equation'', we again find that $\epsilon(t)$
obeys the same equation as the 1D linear growing mode $D_+(t)$, and 
we can write $\epsilon(t) = \epsilon_0 D_+(t)$.

Next, the change of frame described in Eq.(\ref{x-d-v-1}) becomes
\be
x' = (1+\epsilon) x , \;\;\; 
\delta' = \delta - \epsilon (1+\delta) , \;\;\;
v' = v + \dot{\epsilon} a x ,  \;\;\;
\label{x-d-v-1-1D}
\ee
and at linear order over both $\delta$ and $\epsilon$ we have $\delta_L = \delta_L' + \epsilon$.
This means that the background density perturbation
$\epsilon$ is again absorbed by the change of frame, with $\epsilon_0=\Delta\delta_{L0}$.
The redshift-space coordinate $s$ now transforms as
\be
s' = \left( 1 + \epsilon + \frac{\dot\epsilon}{H} \right) s .
\label{sp-s-1D}
\ee
Then, as in Eq.(\ref{map-s-k-1}), the redshift-space density contrast in the actual unprimed
frame, with the uniform overdensity $\Delta\delta_{L0}$, writes as
\be
k \neq 0 : \;\;\; \tdelta^s_{\epsilon_0}(k,t) = 
\tdelta^s[(1-\epsilon-\dot{\epsilon}/H) k,D_{+\epsilon_0},f_{\epsilon_0}] , \;\;\;
\label{deltas-epsilon0-1D}
\ee
where we disregarded the Dirac factor that does not contribute for wave numbers
$k\neq 0$.
Therefore, the derivative of the redshift-space density contrast with respect to $\epsilon_0$
reads as
\be
\frac{\partial \tdelta^s(k,t)}{\partial\epsilon_0} & = & 
\frac{\partial D_{+\epsilon_0}}{\partial\epsilon_0} \frac{\partial \tdelta^s}{\partial D_+} 
+ \frac{\partial f_{\epsilon_0}}{\partial\epsilon_0} \frac{\partial \tdelta^s}{\partial f} \nonumber \\
&& - (1+f) D_+ k \frac{\partial\tdelta^s}{\partial k} .
\label{d-delta-d-eps0-1D-1}
\ee

As shown in \cite{Valageas:2014ab}, the derivative of the linear growing mode is
$\partial D_+/\partial\epsilon_0 = D_+^2$, which means that $D_+'=D_+ + \epsilon_0 D_+^2$.
Then, using $a'=a-\epsilon_0 D_+ a$ and the definition (\ref{eta-1D}) for $f$ and $f'$,
we obtain $f'=f + f (\epsilon+\dot{\epsilon}/H)$, whence
\be
\left. \frac{\partial D_{+\epsilon_0}}{\partial\epsilon_0} \right |_{\epsilon_0=0} = D_+^2 , \;\;\;
\left. \frac{\partial f_{\epsilon_0}}{\partial\epsilon_0} \right |_{\epsilon_0=0} = f D_+ (1+f) . \;\;
\label{d-D+-f-epsilon0-1D}
\ee
Therefore, Eq.(\ref{d-delta-d-eps0-1D-1}) gives Eq.(\ref{d-delta-d-eps0-1D-2}).

\subsection{1D consistency relations}
\label{sec:1D-consistency-relations}

Using the result (\ref{d-delta-d-eps0-1D-2}), the 1D version of the consistency relations
(\ref{tCn0-1}) writes as
\be
\frac{1}{2} \sum_{\pm k'} \langle \tdelta_{L0}(k') \tdelta^s(k_1,t_1) .. 
\tdelta^s(k_n,t_n) \rangle'_{k'\rightarrow 0} =  P_{L0}(k') \nonumber \\
&& \hspace{-7.9cm} \times \sum_{i=1}^n D_{+i} \biggl [ \frac{1+f_i}{n} 
+ \frac{\partial}{\partial\ln D_{+i}} + (1+f_i) f_i \frac{\partial}{\partial f_i} \nonumber \\
&& \hspace{-7.9cm}  - (1+f_i) \sum_{j=1}^n (\delta^K_{i,j} - \frac{1}{n} ) 
k_i \frac{\partial}{\partial k_j} \biggl ]
\langle \tdelta^s(k_1,t_1) .. \tdelta^s(k_n,t_n) \rangle' . \nonumber \\
&& 
\label{tCn0-1-1D}
\ee
The 3D angular average $\int \dd\Omega_{\bfk'}/(4\pi)$ of Eq.(\ref{tCn0-1}) is replaced
by the 1D average $\frac{1}{2} \sum_{\pm k'}$ over the two directions of $k'$
(i.e., the two signs of $k'$). We again defined the reduced polyspectra as in
Eq.(\ref{multi-spectra}), $\langle \tdelta^s(k_1) .. \tdelta^s(k_n) \rangle =
\langle \tdelta^s(k_1) .. \tdelta^s(k_n) \rangle'  \; \delta_D(k_1 \!+\! .. \!+\! k_n)$.

On large scales we recover the linear theory, with 
$\tdelta^s(k',t') \simeq D_+(t') (1+f') \tdelta_{L0}(k')$, and Eq.(\ref{tCn0-1-1D}) also writes as
\be
\frac{1}{2} \sum_{\pm k'} \biggl \langle \frac{\tdelta^s(k',t')}{1+f'} \tdelta^s(k_1,t_1) .. 
\tdelta^s(k_n,t_n) \biggl \rangle'_{k'\rightarrow 0} =  P_L(k',t') \nonumber \\
&& \hspace{-7.9cm} \times \sum_{i=1}^n \frac{D_{+i}}{D_+'} \biggl [ \frac{1+f_i}{n} 
+ \frac{\partial}{\partial\ln D_{+i}} + (1+f_i) f_i \frac{\partial}{\partial f_i} \nonumber \\
&& \hspace{-7.9cm}  - (1+f_i) \sum_{j=1}^n (\delta^K_{i,j} - \frac{1}{n} ) 
k_i \frac{\partial}{\partial k_j} \biggl ]
\langle \tdelta^s(k_1,t_1) .. \tdelta^s(k_n,t_n) \rangle' . \nonumber \\
&& 
\label{tCn-1-1D}
\ee
When all times are equal, $t'=t_1=..=t_n\equiv t$, this simplifies as Eq.(\ref{tCn-2-1D}).

\subsection{Zel'dovich solution}
\label{sec:Zel'dovich}

In the 1D case, the Zel'dovich approximation is exact until shell crossing
\cite{ZelDovich1970,Valageas:2014ab}.
It corresponds to taking for the particle trajectories the linear prediction,
 \be
x(q,t) = q + \Psi_L(q,t) 
\label{Zel-def}
\ee
with
\be
\Psi_L(q)= \ii \int_{-\infty}^{+\infty} \frac{\dd k}{k} \; e^{\ii k q} \; \tdelta_L(k,t) .
\label{Psi-tdL}
\ee
Therefore, the redshift-space coordinate (\ref{s-def-1D}) writes as (using $v=a\dot{x}$)
\be
s = q + (1+f) \Psi_L ,
\label{s-Zel}
\ee
and the redshift-space nonlinear density contrast (\ref{tdelta-s-def-1D}) as 
Eq.(\ref{tdelta-s-Zel}).

\subsection{Check of the 1D consistency relations}
\label{sec:1D-check}

\subsubsection{Impact of a large-scale perturbation on the nonlinear redshift-space density contrast}
\label{sec:Impact-1D}

To check the validity of the 1D consistency relations from the exact solution
(\ref{tdelta-s-Zel}), we simply need the change of the nonlinear redshift-space density contrast
$\tdelta^s(k)$ when we make a small perturbation $\Delta\delta_{L0}$ to the initial
conditions on much larger scales.
Let us consider the impact of a small large-scale perturbation $\Delta \delta_{L0}$ to the
initial conditions. Here we also restrict to even perturbations, 
$\Delta\tdelta_{L0}(-k') =  \Delta\tdelta_{L0}(k')$, as the consistency relations studied in this
paper apply to spherically averaged statistics, which correspond to the $\pm k'$
averages in the 1D relations (\ref{tCn0-1-1D})-(\ref{tCn-1-1D}).
Then, expanding Eq.(\ref{tdelta-s-Zel}) up to first order over $\Delta \delta_{L0}$, and over
powers of $k'$, we obtain
\be
\hspace{-0.5cm} k' \rightarrow 0 : \;\;\; \Delta\tdelta^s(k) & = & 
(1+f) D_+ \left[ \int \dd k' \Delta\tdelta_{L0}(k') \right] \nonumber \\
&& \hspace{-2.7cm} \times \int \frac{\dd q}{2\pi} \, e^{-\ii k q + k (1+f) D_+ \int \frac{\dd k''}{k''} 
e^{\ii k'' q} \tdelta_{L0}(k'') }  (\ii k q) . \;\;
\label{Delta-Z-1}
\ee
Here the limit $k'\rightarrow 0$ means that we consider a perturbation of the initial
conditions $\Delta\tdelta_{L0}(k')$ that is restricted to low wave numbers, $k'<\Lambda$,
with a cutoff $\Lambda$ that goes to zero (i.e., that is much smaller than the wave
numbers $k$ and $2\pi/q$ of interest).

On the other hand, from the expression (\ref{tdelta-s-Zel}) we obtain at once the exact
result
\be
\hspace{-0cm} \frac{\partial\tdelta^s}{\partial\ln D_+} + (1+f) f \frac{\partial\tdelta^s}{\partial f} 
- (1+f) k \frac{\partial\tdelta^s}{\partial k} & = &  \int \frac{\dd q}{2\pi} \, 
e^{-\ii k q} \nonumber \\
&& \hspace{-5.5cm} \times \; e^{k(1+f) D_+\int \frac{\dd k''}{k''} e^{\ii k'' q} \tdelta_{L0}(k'') }  
(1+f) (\ii k q) . \;\;
\label{tdelta-Z-2}
\ee
The comparison with Eq.(\ref{Delta-Z-1}) gives
\be
\hspace{-0cm} k' \rightarrow 0 : \;\; \Delta\tdelta^s(k) = D_+ \left[ \int \dd k' \Delta\tdelta_{L0}(k') 
\right]  \left( \frac{\partial\tdelta^s(k)}{\partial\ln D_+} \right. \nonumber \\
&& \hspace{-6.7cm} \left. + (1+f) f \frac{\partial\tdelta^s(k)}{\partial f}  
- (1+f) k \frac{\partial\tdelta^s(k)}{\partial k} \! \right) . \;\;
\label{Delta-Z-2}
\ee

The consistency relations (\ref{tCn0-1-1D})-(\ref{tCn-1-1D}) and (\ref{tCn-2-1D}) only rely
on the expression (\ref{d-delta-d-eps0-1D-2}), which also reads (at linear order over
$\epsilon_0$) as
\be
\hspace{-0cm} \Delta\tdelta^s(k) = \epsilon_0 \, D_+ \, 
\left( \frac{\partial\tdelta^s(k)}{\partial\ln D_+} + (1+f) f \frac{\partial\tdelta^s(k)}{\partial f}  \right.
\nonumber \\
&& \hspace{-5cm} \left. - (1+f) k \frac{\partial\tdelta^s(k)}{\partial k} \right) .
\label{Delta-Z-3}
\ee
Since we have $\epsilon_0 = \Delta \delta_{L0} = \int \dd k' \Delta\tdelta_{L0}(k')$, we recover 
Eq.(\ref{Delta-Z-2}).
This provides an explicit check of Eq.(\ref{d-delta-d-eps0-1D-2}), hence of the 1D consistency
relations.

\subsubsection{Explicit check on the redshift-space density polyspectra}
\label{sec:Explicit-1D}

Instead of looking for the impact of a large-scale linear perturbation on the nonlinear
density contrast, as in Sec.~\ref{sec:Impact-1D}, we can directly check the consistency
relations in their forms (\ref{tCn0-1-1D}) or (\ref{tCn-2-1D}).
Considering for simplicity the equal-time polyspectra (\ref{tCn-2-1D}), we define
the mixed polyspectra, formed by one linear density contrast and $n$ nonlinear redshift-space
density contrasts,
\be
\hspace{-0.5cm} E_n^s(k';k_1,..,k_n;t) & \equiv &
\langle \tdelta_L(k',t) \tdelta^s(k_1,t) .. \tdelta^s(k_n,t) \rangle \nonumber \\
&& \hspace{-2.3cm} = D_+ \biggl \langle \tdelta_{L0}(k') \int 
\frac{\dd q_1 .. \dd q_n}{(2\pi)^n} \; e^{-\ii \sum_{j=1}^n  k _j  \, q_j} \nonumber \\
&& \hspace{-2cm} \times \; e^{(1+f) D_+ \int \dd k/k \; \tdelta_{L0}(k) 
\sum_{j=1}^n k_j \,  e^{\ii k q_j} } \biggl \rangle ,
\ee
where in the last expression we used Eq.(\ref{tdelta-s-Zel}).
The Gaussian average over the initial conditions $\tdelta_{L0}$ gives
\be
E_n^s  & = & - \frac{P_L(k')}{k'} (1+f) \int \frac{\dd q_1 .. \dd q_n}{(2\pi)^n} \; 
\sum_{j=1}^n k_j \, e^{-\ii k' q_j} \nonumber \\
&& \hspace{-0.7cm} \times \; e^{-\ii \sum_{j=1}^n k_j \, q_j
-(1+f)^2 D_+^2/2 \int  \dd k/k^2 \; P_{L0}(k) 
\left| \sum_{j=1}^n k_j e^{\ii k q_j} \right|^2 } . \nonumber \\
&&
\ee
Making the changes of variable $q_1=q_1'+q_n$, .., $q_{n-1}=q_{n-1}'+q_n$,
the argument of the last exponential does not depend on $q_n$.
Then, the integration over $q_n$ yields a Dirac factor $\delta_D(k'+k_1+..+k_n)$,
that we factor out by defining $E_n^s = E_n^{s'} \delta_D(k'+k_1+..+k_n)$,
with a primed notation as in Eq.(\ref{multi-spectra}), and we replace $k_n$
by $-(k'+k_1+..+k_{n-1})$.
Finally, in the limit $k' \rightarrow 0$ we expand the terms $e^{-\ii k' q_j}$ up to first
order over $k'$, and we obtain
\be
\hspace{-0.5cm} k' \rightarrow 0 & : & \;\; E_n^{s'} = P_L(k') (1+f) \int \frac{\dd q_1 .. \dd q_{n-1}}{(2\pi)^{n-1}}
\nonumber \\
&& \hspace{-1.1cm} \times \; \biggl [ 1 + \ii \sum_{j=1}^{n-1} k_j \, q_j \biggl ] 
\; e^{-\ii \sum_{j=1}^{n-1} k_j \, q_j } \nonumber \\
&& \hspace{-1.1cm} \times \; e^{-(1+f)^2 D_+^2/2 \int \dd k/k^2 \; P_{L0}(k) 
\left|  \sum_{j=1}^{n-1} k_j ( e^{\ii k q_j}-1) \right|^2 } . \;\;
\label{En-1D-1}
\ee

Proceeding in the same fashion, the $n-$point redshift-space polyspectra read as
\be
\hspace{-0.6cm} P_n^s \equiv \langle \tdelta^s(k_1,t) .. \tdelta^s(k_n,t) \rangle' &&  \nonumber \\
&& \hspace{-3.9cm} = \int \frac{\dd q_1 .. \dd q_{n-1}}{(2\pi)^{n-1}}
\; e^{-\ii \sum_{j=1}^{n-1} k_j \, q_j} \nonumber \\
&& \hspace{-3.9cm} \times \; e^{-(1+f)^2 D_+^2/2 \int \dd k/k^2 \; P_{L0}(k) 
\left|  \sum_{j=1}^{n-1} k_j ( e^{\ii k q_j}-1) \right|^2 } . 
\label{Pn-1D}
\ee
Then, we can explicitly check from the comparison with Eq.(\ref{En-1D-1}) that
we have the relation
\be
k' \rightarrow 0 : \;\; E_n^{s'} & = & P_L(k') \left[ 1 + f + \frac{\partial}{\partial\ln D_+} 
+ (1+f) f \frac{\partial}{\partial f} \right. \nonumber \\
&&  \left. - (1+f) \sum_{i=1}^{n-1} \frac{\partial}{\partial\ln k_i} \right] P_n^s ,
\label{En-Pn}
\ee
and we recover the consistency relation (\ref{tCn-2-1D}). [In Eq.(\ref{En-Pn}) the
right-hand side does not involve $k_n$ because it has been replaced by
$-(k_1+..+k_{n-1})$ in Eq.(\ref{Pn-1D}), using the Dirac factor $\delta_D(k_1+..+k_n)$.]

\bibliography{ref1}

\end{document}